\documentclass{aastex61}
\usepackage{multirow}
\usepackage{captcont}

\received{June, 2020}
\revised{Sept, 2020}
\accepted{Oct, 2020}
\submitjournal{AJ}

\shorttitle{Diversity of asymmetries}
\shortauthors{van der Marel et al.}

\begin{document}

\title{On the diversity of asymmetries in gapped protoplanetary disks}

\correspondingauthor{Nienke van der Marel}
\email{astro@nienkevandermarel.com}

\author{Nienke van der Marel}
\affil{Physics \& Astronomy Department, 
University of Victoria, 
3800 Finnerty Road, 
Victoria, BC, V8P 5C2, 
Canada}
\affil{Banting Research fellow}

\author{Til Birnstiel}
\affil{University Observatory, Faculty of Physics, Ludwig-Maximilians-Universität München, Scheinerstr. 1, 81679 Munich, Germany}
\affil{Exzellenzcluster ORIGINS, Boltzmannstr. 2, D-85748 Garching, Germany}

\author{Antonio Garufi}
\affil{INAF, Osservatorio Astrofisico di Arcetri, Largo Enrico Fermi 5, 50125, Firenze, Italy}

\author{Enrico Ragusa}
\affil{School of Physics and Astronomy, University of Leicester, Leicester, United Kingdom}

\author{Valentin Christiaens}
\affil{School of Physics and Astronomy, Monash University, Vic 3800, Australia}

\author{Daniel J. Price}
\affil{School of Physics and Astronomy, Monash University, Vic 3800, Australia}

\author{Steph Sallum}
\affil{Department of Physics and Astronomy, 4129 Frederick Reines Hall, University of California, Irvine, CA, 92697-4575, USA}

\author{Dhruv Muley}
\affil{Physics \& Astronomy Department, 
University of Victoria, 
3800 Finnerty Road, 
Victoria, BC, V8P 5C2, 
Canada}

\author{Logan Francis}
\affil{Physics \& Astronomy Department, 
University of Victoria, 
3800 Finnerty Road, 
Victoria, BC, V8P 5C2, 
Canada}

\author{Ruobing Dong}
\affil{Physics \& Astronomy Department, 
University of Victoria, 
3800 Finnerty Road, 
Victoria, BC, V8P 5C2, 
Canada}

\begin{abstract}
Protoplanetary disks with large inner dust cavities are thought to host massive planetary or substellar companions. These disks show asymmetries and rings in the millimeter continuum, caused by dust trapping in pressure bumps, and potentially vortices or horseshoes. The origin of the asymmetries and their diversity remains unclear. We present a comprehensive study of 16 disks for which the gas surface density profile has been constrained by CO isotopologue data. We compare the azimuthal extents of the dust continuum profiles with the local gas surface density in each disk, and find that the asymmetries correspond to higher Stokes numbers or low gas surface density. We discuss which asymmetric structures can be explained by a horseshoe, a vortex or spiral density waves. 
Second, we reassess the gas gap radii from the $^{13}$CO maps, which are about a factor 2 smaller than the dust ring radii, suggesting that companions in these disks are in the brown dwarf mass regime ($\sim 15-50 M_{\rm Jup}$) or in the Super-Jovian mass regime ($\sim 3-15 M_{\rm Jup}$) on eccentric orbits. This is consistent with the estimates from contrast curves on companion mass limits. These curves rule out (sub)stellar companions ($q>$0.05) for the majority of the sample at the gap location, but it remains possible at even smaller radii. 
Third, we find that spiral arms in scattered light images are primarily detected around high luminosity stars with disks with wide gaps, which can be understood by the dependence of the spiral arm pitch angle on disk temperature and companion mass.
\end{abstract}


\section{Introduction}
Protoplanetary disks around young stars are the birth places of planets, and their observed structures reveal the result of planet-disk interactions. Of particular interest are the so-called transition disks with large inner dust cavities ($>$20 au), \citep[e.g.][]{Espaillat2014,vanderMarel2017}. In this work, we use the term transition disk for any disk with a large cleared inner dust cavity ($>$20 au) as revealed by millimeter observations. ALMA has revealed a large diversity of structures in transition disks in both the dust \citep[e.g.][]{Pinilla2018tds,vanderMarel2019},  and the gas \citep[e.g.][]{vanderMarel2016-isot,Dong2017,Boehler2017},  showing deep gas cavities well within the dust ring radii. The presence of these gas cavities is consistent with clearing by massive companions (either planetary or substellar) at wide orbits where the millimeter dust is trapped at the edge of the gap \citep[e.g.][]{Pinilla2012b}. Another proposed scenario for transition disk cavities is photoevaporation \citep{Alexander2014}, which is generally ruled out by the high accretion rates \citep{OwenClarke2012}. Also dead zones (low viscosity regions due to poor ionization) have been proposed to generate transition disk cavities due to their sharp viscosity gradient \citep[][]{Regaly2012} but the deep observed gas gaps cannot be reproduced by dead zones alone \citep{Pinilla2016dz}.

Some dust rings are highly asymmetric at millimeter wavelengths thought to be caused by azimuthal trapping \citep[e.g.][]{vanderMarel2013,Birnstiel2013}, but only a fraction of the transition disks is asymmetric. For a large sample of 38 transition disks, which are all known transition disks resolved at high spatial resolution with ALMA, the fraction of asymmetric disks is 24\% \citep{Francis2020}, but the completeness with respect to the total disk population cannot be determined. Regardless of the exact fraction, it remains unclear why azimuthal trapping only occurs in some of these disks.

The two main origins of azimuthal dust traps (azimuthal gas pressure maxima) are long-lived anticyclonic vortices, caused by the Rossby Wave Instability at the outer edge of the companion gap \citep[e.g.][]{BargeSommeria1995,Zhu2014} and gas horseshoes due to a pile-up of material in eccentric cavities due to a binary companion \citep[e.g.][]{Ragusa2017}, with a mass ratio requirement of $q>0.05$. Whereas long-lived vortices require a low viscosity in the disk ($\alpha\leq10^{-4}$) to survive \citep{Godon1999,Regaly2012}, horseshoes do not dissipate even at high viscosity \citep{Miranda2017,Ragusa2020}. Both scenarios produce an azimuthal gas overdensity of a factor of $\lesssim$2, which can trap millimeter-sized dust efficiently in radial and azimuthal direction, resulting in a significant dust asymmetry \citep{Birnstiel2013}. Trapping efficiency increases with grain size up to a Stokes number of 1 \citep{Birnstiel2013,Birnstiel2016}. The Stokes number $St$ is defined (see Eqn. \ref{eqn:stokes} below) as the stopping time of a dust particle per orbital time and indicates how well dust grains are coupled to the gas. The gas overdensity itself co-moves with the gas on a Keplerian orbit. A third possibility for a dust asymmetry is an eccentric disk. In contrast to a vortex or horseshoe, an eccentric disk caused by a massive companion \citep{KleyDirksen2006} does not co-move with the gas \citep{Ataiee2013} and thus does not trap millimeter dust: it acts as a 'traffic jam' in their apocenter. The observed segregation between gas and dust consistent with trapping already rules out eccentricity as a major explanation for most observed asymmetric dust disks to date \citep{Ataiee2013,vanderMarel2016-isot}. The differences between these three types of disks is summarized in Table \ref{tbl:scenarios}.

\begin{table}[!ht]
\centering
\caption{Possible scenarios for an asymmetric disk caused by a companion}
\label{tbl:scenarios}
\begin{tabular}{lllll}
\hline
Scenario & Required $q$ companion & Required $\alpha$ & Co-moving/trapping? & Ref\\
\hline
Vortex & $>$0.0002\footnote{Based on a minimum of 68 $M_{\rm Earth}$. The actual minimum mass estimate depends on the value of $\alpha$: the listed value is derived for $\alpha\sim10^{-4}$, as in \citet{Dong2018gaps}.} & $\lesssim10^{-4}$&Y&\citet{Zhu2014,Dong2018gaps} \\
Horseshoe & $>$0.05 & any & Y & \citet{Ragusa2017} \\
Eccentric disk & $0.003-0.05$ & any & N & \citet{Ataiee2013} \\
\hline
\end{tabular}
\end{table}

Vortex dissipation due to dust feedback \citep[e.g.][]{Fu2014,Miranda2017} and slowly growing planets \citep{Hammer2017} could potentially limit the vortex lifetime. Gas horseshoes are expected to survive for very long timescales ($>$7000 orbits), consistent with the disk lifetime \citep{Miranda2017,Ragusa2020}. A dissipation process would be a possible explanation for the low occurrence rate and diversity of asymmetries in transition disk rings, but this has not been quantified.

The main observable distinction between the vortex and gas horseshoe mechanisms is the companion mass and location: gas horseshoes require a mass ratio $q>0.05$ (implying substellar rather than planetary mass) and the companion is closer to the star, compared to the radial dust asymmetry location. HD142527 has been shown to host a (sub)stellar M-dwarf companion with $M\sim0.26M_{\odot}$ at an eccentric orbit between 18-57 au \citep{Lacour2016,Claudi2019}. Therefore, a horseshoe has been invoked to explain the asymmetry in the HD142527 disk \citep{Price2018}. For most transition disks, it is unknown whether a companion, either planetary or (sub)stellar, is present inside the disk, in particular in the inner part.

The detection and quantification of companions in transition disks through direct imaging remains challenging, due to the high contrast required to detect a companion in a dusty environment. Companion candidates have been debated in e.g. HD169142 \citep{Biller2014,Ligi2018}, LkCa15 \citep{Sallum2015,Thalmann2016,Currie2019}, HD~100546 \citep{Quanz2013,Currie2015,Rameau2017} and MWC758 \citep{Reggiani2018,Wagner2019}. The only robust detections of planetary companions in a transition disk to date are PDS70b and c \citep{Keppler2018, Haffert2019}. 
In most transition disks no detections of companions have been found, and only upper limits have been derived. The low number of companion detections in transition disks has been suggested to be caused by uncertainties in expected planet brightness. If young planets are faint, they might only be detectable during the initial accretion phase while material is still flowing through the gap \citep{Francis2020} or during an episodic accretion outburst \citep{Brittain2020}. Also, at distances close to the star ($<$0.15") the achievable contrast remains limited. 

Indirect evidence for companions is found in wide, deep gas gaps observed through CO isotopologues observations in transition disks \citep[e.g.][]{vanderMarel2016-isot}. The deep density drops and large separation between the gas cavity radius and dust ring radius already suggests that massive companions ($>5 M_{\rm Jup}$) must be responsible for the gaps \citep{vanderMarel2016-isot,Facchini2017gaps}. Eccentric companions have been suggested to explain the wide separation between dust and gas cavity radius \citep{Muley2019}. CO isotopologue images reveal a complete deficit of material close to the star in many transition disks \citep{vanderMarel2016-isot}, which is generally modeled using a prescription of the surface density with a cavity, depleted of gas all the way down to the centre of the disk. This parametrization is inconsistent with the morphology of planet-induced gaps in planet-disk interaction models, which generally show a gap around the planet orbit but an undisturbed gas surface density profile inside the planet orbit \citep[e.g][]{Fung2016,Facchini2017gaps}. 
A fully cleared gas cavity would be more consistent with a more massive stellar companion, such as suggested for HD142527 \citep{Price2018} or perhaps multiple planets. However, for most CO observations of transition disks the amount of gas inside the cavity (in particular close to the star) cannot be constrained due to the low spatial resolution, typically 0.25" or $\sim$35 au, blending the contributions of the outer and inner gap edge, and a distinction between gas gap and cavity cannot be made \citep{vanderMarel2018a}. 

Other indirect evidence for companions is found through CO kinematics in disks of non-Keplerian motion: so-called `kinks' in the channel maps in between dust rings due to spiral density waves launched by the companion \citep[e.g.][]{Pinte2018,Pinte2019}, pressure perturbations \citep{Teague2018}, meridional flows \citep{Teague2019} and warps in the inner cavity of the disk \citep[e.g.][]{Casassus2013,Boehler2017,Mayama2018}. These warps can be explained by either misaligned inner disks or fast radial flows \citep{Rosenfeld2014,Facchini2017warps, Zhu2019}, although the reason could also be natal disk structure \citep{Bate2018}. Also spiral arms seen in scattered light images have been linked to the presence of companions \citep{Dong2015spirals} and are often found in asymmetric disks \citep{Garufi2018}. Asymmetries have been proposed to trigger spiral arms  \citep{vanderMarel2016-spirals,Cazzoletti2018}, or be part of it \citep{Dong2018,Rosotti2019}, but there is no universal explanation for their coappearance.

\begin{figure}[!ht]
\centering
\includegraphics[width=0.9\textwidth]{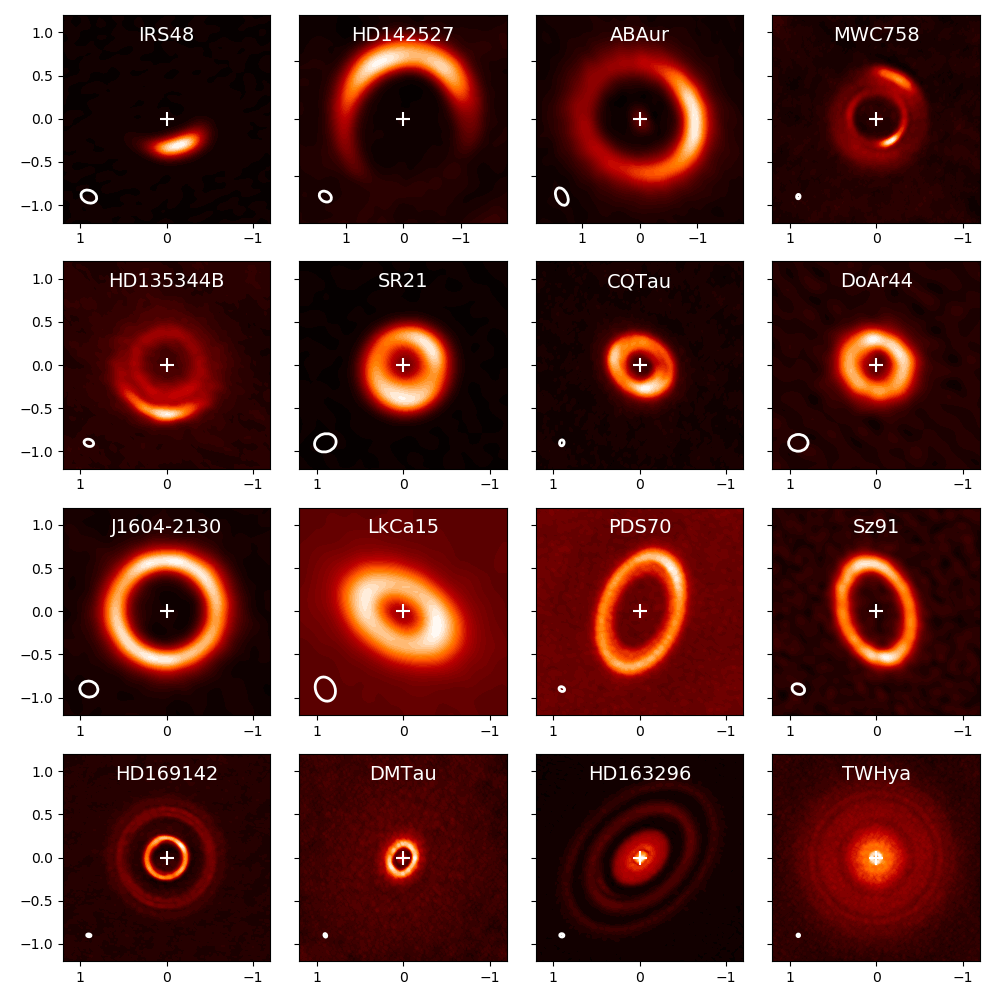}
\caption{Gallery of ALMA continuum images at Band 6 or 7 of the disks in the sample of this study. The maps of HD~142527 and AB~Aur are zoomed out compared to the others because of their size. Details of the observations can be found in Table \ref{tbl:sample}.}
\label{fig:gallery}
\end{figure}

With the large number of observed morphologies of both asymmetries, rings and spiral arms it remains unclear how these different structures are connected to each other, and whether the diversity is related to evolutionary, dynamical or stellar effects. In order to understand the diversity in disk structures, we present a sample study of 16 disks with and without asymmetries for a range of stellar, disk and companion properties. We analyze the gas gap properties from spatially resolved CO observations and the Stokes numbers of dust grains throughout the disk and compare these with the dust properties. Furthermore, we compare the disk profiles with the limits of companion studies.

The paper is structured as follows. In Section \ref{sct:sample} we present the sample selection, based on a number of criteria and required data. We derive azimuthal profiles in the dust rings in Section \ref{sct:asym} from ALMA archival continuum data, gas gaps from ALMA CO observations in Section \ref{sct:gas} and what is known about companions from direct imaging in Section \ref{sct:comps}. Using the combined information of gas and dust we analyze in Section \ref{sct:analysis} the gas surface density profiles, the relevant Stokes numbers in asymmetries and properties of spiral arms. In Section \ref{sct:discussion} and \ref{sct:conclusions} we discuss the implications of our analysis and summarize our main conclusions.

\begin{table*}[!ht]
\caption{Sample properties}
\label{tbl:sample}
\begin{tabular}{lllllllllllllll}
\hline
Target & M & $\lambda_{obs}$ & Program ID& $R_{dust}$ &FWHM&$R_{ind}$& PA & $i$& $d$ & SpT & $M_*$& S& C& Refs\\
&&(mm)&&(au)&($^{\circ}$)&(au)&($^{\circ}$)&($^{\circ}$)&(pc)&&($M_{\odot}$)&&&\\
&(1)&(2)&&(3)&(4)&(5)&&&&&(6)&(7)&(8)&(9)\\
\hline
IRS48	&	A	&	0.9	&	2013.1.00100.S	&	70	&	58&- &	100	&	50	&	134	&	A0	&	2.0	&	Y	&	-	&	1,2,-	\\
HD142527	&	A	&	0.9	&	2012.1.00631.S	&	180	&	155& 4.1 &	-20	&	27	&	157	&	F6	&	1.7	&	Y	&	Y	&	1,3,4	\\
ABAur	&	A	&	0.9	&	2012.1.00303.S	&	170	&	122&4.7 &	50	&	23	&	163	&	A0	&	2.6	&	Y	&	-	&	1,5,-	\\
MWC758	&	A	&	0.9	&	2017.1.00492.S	&	50	&49&	3.0	&	62	&	21	&	160	&	A8	&	1.7	&	Y	&	C	&	6,7,8	\\
&&&&90&47&&&&&&&&&\\
HD135344B	&	A	&	1.9	&	2016.1.00340.S	&	51&	-&- & 62	&	18	&	135	&	F4	&	1.4	&	Y	&	L	&	9,10,11	\\
&&&&79&96&&&&&&&&&\\
SR21	&	A	&	0.9	&	2012.1.00158.S	&	36	&- &	- &	24	&	15	&	138	&	G3	&	2.1	&	Y	&	L	&	12,13,13	\\
&&&&55&82&&&&&&&&&\\
&&&&58&165&&&&&&&&&\\
CQTau	&	A	&	1.3	&	2017.1.01404.S	&	50	&59&- & 55	&	35	&	162	&	F2	&	1.7	&	Y	&	L	&	14,15,15	\\
&&&&50&59&&&&&&&&&\\
DoAr44	&	S	&	0.9	&	2012.1.00158.S	&	47	&-& - &	60	&	20	&	146	&	K3	&	1.4	&	N	&	-	&	12,16,-	\\
J1604-2130	&	S	&	0.9	&	2015.1.00888.S	&	85	&-& - &	80	&	6	&	150	&	K2	&	1.0	&	N	&	L	&	17,18,19	\\
LkCa15	&	S	&	1.1	&	2015.1.00678.S	&	75	&-&	- & 60	&	55	&	159	&	K2	&	1.3	&	N	&	C	&	20,21,22	\\
PDS70	&	S	&	0.9	&	2017.A.00006.S	&	74	&-& 10 &	-20	&	52	&	113	&	K7	&	0.9	&	N	&	Y	&	23,24,25	\\
Sz91	&	S	&	0.9	&	2012.1.00761.S	&	94	&-&	- & 17	&	45	&	159	&	M1	&	0.6	&	N	&	L	&	26,27,27	\\
HD169142	&	S	&	1.3	&	2016.1.00344.S	&	25	& -&  2.2 &	5	&	13	&	114	&	F1	&	1.7	&	Y?	&	C	&	28,29,30	\\
&&&&60&-&&&&&&&&&\\
DMTau	&	S	&	1.3	&	2017.1.01460.S	&	25	&- &	7.5 & 158	&	35	&	145	&	M2	&	0.5	&	-	&	-	&	31,-,-	\\
HD163296	&	S	&	1.3	&	2016.1.00484.L	&	14,67,100	&- &	- & 132	&	42	&	102	&	A1	&	2.0	&	N	&	L	&	32,33,34	\\
TWHya	&	S	&	0.9	&	2015.1.00686.S	&	12,29,40	&-& 1.0 &	-25	&	6	&	60	&	M1	&	0.4	&	N	&	L	&	35,36,37	\\
\hline
\end{tabular}
Explanation columns. (1) Dust morphology: A=asymmetric, S=symmetric; (2) Observing wavelength of the used dust continuum observations; (3) Peak radius of the dust ring(s) in the disk; (4) Azimuthal extent of the asymmetric feature along the dust ring from visibility analysis; (5) Size of the inner dust disk, taken from \citet{Francis2020}; (6) Stellar masses have been taken from \citet{Francis2020}, who derived them using \textit{Gaia DR2} distances and \citet{Baraffe2015} and \citet{Siess2000} evolutionary models; (7) Detected spirals in scattered light observations; (8) Companions from direct imaging: Y=companion confirmed, C=companion candidate, L=no detection but limits available; (9) References for respectively ALMA data, spiral arms, companions:
 1) \citet{Francis2020}, 2) \citet{Follette2015}, 3) \citet{Avenhaus2014}, 4) \citet{Claudi2019}, 5) \citet{Boccaletti2020}, 6) \citet{Dong2018}, 7) \citet{Benisty2015}, 8) \citet{Reggiani2018}, 9) \citet{Cazzoletti2018}, 10) \citet{Stolker2016}, 11) \citet{Maire2017}, 12) \citet{vanderMarel2016-isot}, 13) \citet{Muro-Arena2020}, 14) \citet{Ubeira2019}, 15) \citet{Uyama2019}, 16) \citet{Avenhaus2018}, 17) \citet{Mayama2018}, 18) \citet{Mayama2012}, 19) \citet{Canovas2017}, 20) \citet{Qi2019}, 21) \citet{Thalmann2016}, 22) \citet{Thalmann2010}, 23) \citet{Keppler2019}, 24) \citet{Muller2018}, 25) \citet{Mesa2019pds}, 26) \citet{Tsukagoshi2019}, 27) \citet{Mauco2019}, 28) \citet{Perez2019}, 29) \citet{Gratton2019}, 30) \citet{Reggiani2014}, 31) \citet{Kudo2018}, 32) \citet{Huang2018}, 33) \citet{Muro-Arena2018}, 34) \citet{Mesa2019hd16}, 35) \citet{Andrews2016}, 36) \citet{vanBoekel2016}, 37) \citet{Ruane2017}
\end{table*}

\section{Sample}
\label{sct:sample}

We select a sample of disks with gaps, primarily transition disks, in order to constrain dust and gas properties across a range of azimuthal contrast and presence of companions and spiral arms. In order to study the coupling of the dust to the gas, the measurement of the Stokes number is required, as larger Stokes numbers imply decoupling from the gas. The Stokes number is defined as the stopping time of a dust particle divided by the orbital time \citep{Birnstiel2010}. In the Epstein regime (where the ratio of the mean free path of the gas molecules $\lambda_{mfp}$ to the grain size $a_{grain}$ satisfies $\lambda_{mfp}/a_{grain} \geq 4/9$), the Stokes number is defined as 
\begin{equation}
\label{eqn:stokes}
St = \frac{a_{grain}\rho_s\pi}{2\Sigma_{\rm gas}}
\end{equation}
with $\rho_s$ the intrinsic dust density (taken as 1 g cm$^{-3}$) and $\Sigma_{\rm gas}$ the local gas surface density.

We thus require measurements of the gas surface density as function of position in the disk. Gas surface density profiles can be derived from spatially resolved CO isotopologue data in combination with physical-chemical modeling, in order to take into account freeze-out, (isotope-selective) photodissociation, and heating-cooling effects throughout the disk, e.g. DALI \citep{Bruderer2012,Bruderer2013}. Parametric approaches of the abundance and temperature can be informative as well, in particular when multiple CO transitions are used, but larger uncertainties in the derived surface density profile remain. 

We select transition disks for which the gas surface density has been derived using detailed CO modeling and resolved CO isotopologue observations, preferably a combination of $^{13}$CO and optically thin C$^{18}$O. Furthermore, we require that the dust rings are at least marginally spatially resolved in the radial direction, in order to make a proper assessment of the azimuthal structure in the dust. 

The final sample thus consists of 14 transition disks (see Table \ref{tbl:sample}) six of which show asymmetric features, and 2 ring disks without a large inner dust cavity. These two ring disks, TW Hya and HD163296, were added to the sample for comparison, as dust rings in `full' protoplanetary disks are thought to behave in a similar way as transition disk rings with regard to dust trapping \citep{vanderMarel2019}. Some known asymmetric transition disks had to be omitted due to lack of high resolution gas observations \citep[e.g. V1247 Ori and HD143006, ][]{Kraus2017,Andrews2018dsharp} whereas others did not have high resolution dust continuum data \citep[e.g. Sz111, RY Lup, LkH$\alpha$330, ][]{vanderMarel2018a,Isella2013}. The sample covers a range of stellar properties, with spectral types ranging from A0 to M2. Distances are taken from the \emph{Gaia DR2} \citep{Gaia2018}. References for stellar properties are given in \citet{Francis2020}, where the stellar masses have been rederived using the updated \textit{Gaia DR2} distances.

For most of these disks, multiple ALMA programs are available in the ALMA archive. The programs with the best combination of spatial resolution and signal-to-noise ratio are chosen for this study (see Table \ref{tbl:sample}). The ALMA data reduction is described in \citet{Francis2020} for the majority of the disks and the Table lists the reference where the data were first presented. For HD163296, we use the fits file provided by the DSHARP team \citep{Andrews2018dsharp} and for TW Hya the fits file from \citet{Andrews2016}. Table \ref{tbl:sample} also lists the derived radii of the dust inner disk from \citet{Francis2020}.

\section{Data}
\subsection{Dust structure}
\label{sct:asym}

\begin{figure}[!ht]
\includegraphics[width=0.45\textwidth,trim=0 50 0 0]{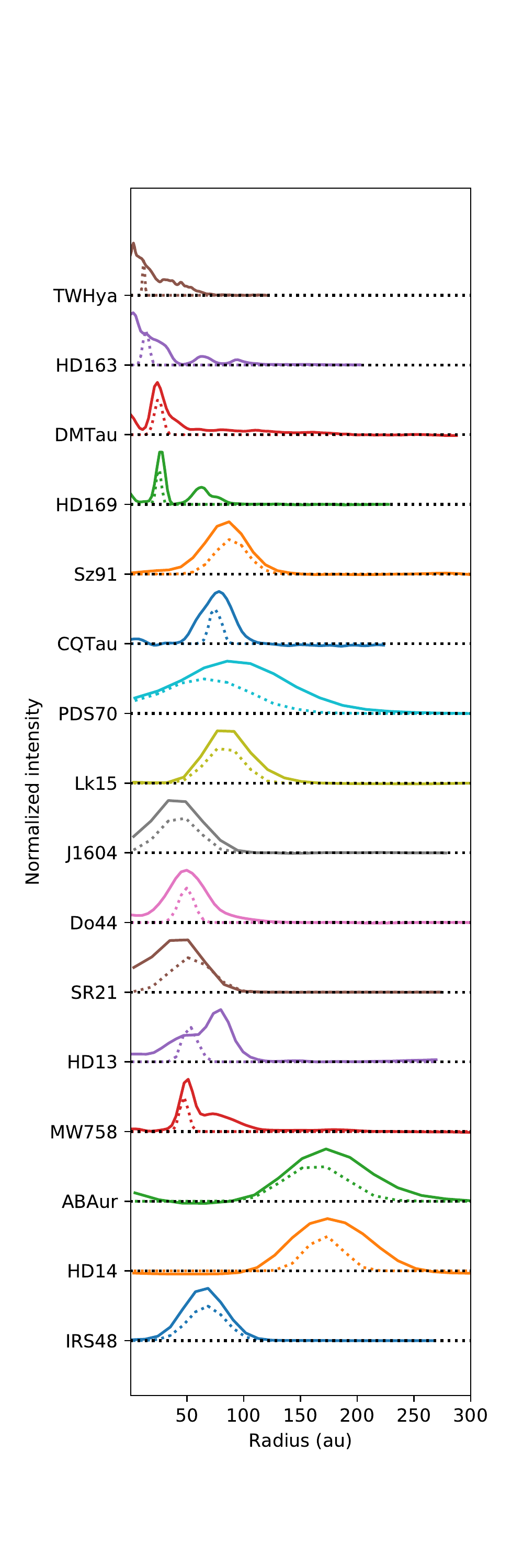}
\includegraphics[width=0.45\textwidth]{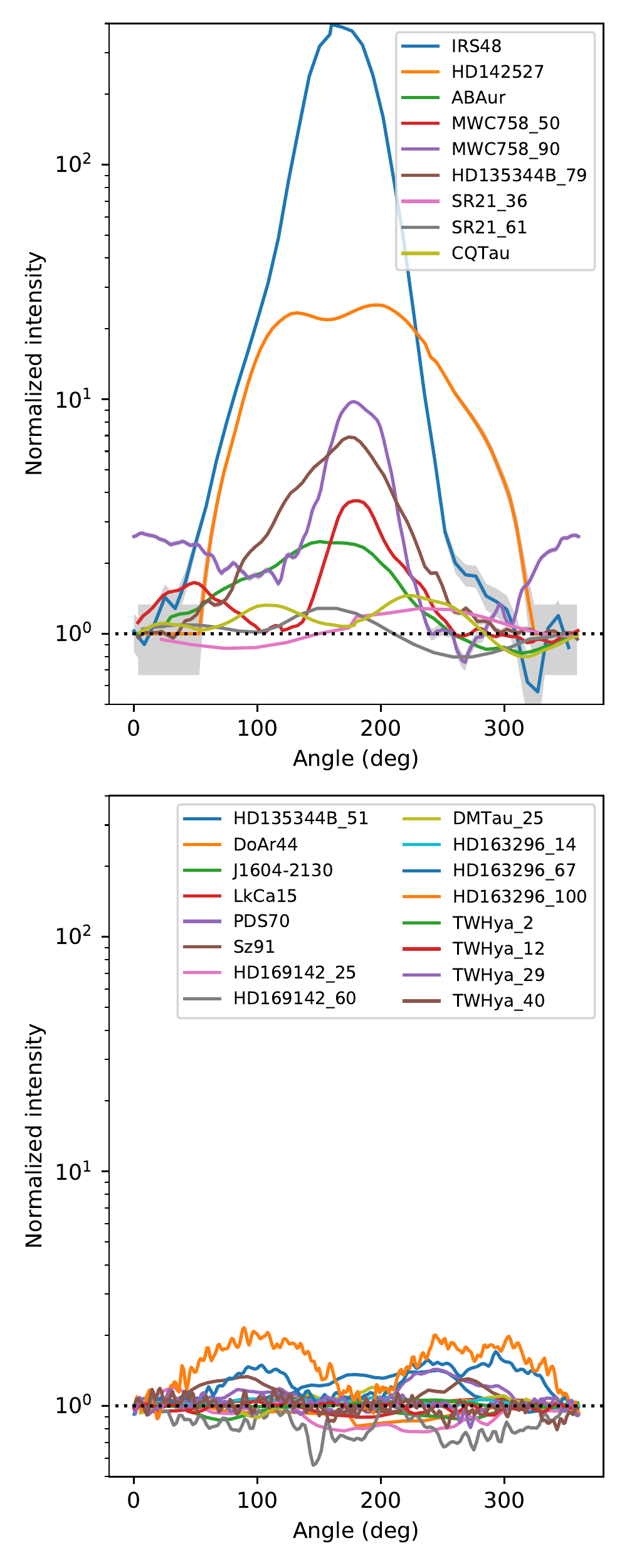}
\caption{Intensity profiles along the radial (left) and azimuthal (right) directions at the location of the dust ring/asymmetry, normalized to the intensity at the opposite side of the asymmetry. The grey areas in the azimuthal plots indicate the noise level. In the left panel, the intensity profile is solid, while the average radial beam size is indicated with a dashed profile. The numbers in the azimuthal curves indicate the radius of the corresponding dust ring.}
\label{fig:azimprofiles}
\end{figure}

The ALMA continuum images of the samples are presented in Figure \ref{fig:gallery}. All images have a high signal-to-noise ratio (SNR): the median SNR of these images is 55, and the lowest SNR is 40. The azimuthal and radial profiles are extracted for each image, using the position angle and inclination of the outer disk.
As most asymmetric disks have a moderate to face-on inclination, optical depth effects are not considered to be affecting the results significantly. The only exception is IRS~48 with a 50$^{\circ}$ inclination, which implies that the continuum contrast might be overestimated by a factor of a few. The radial profiles are taken by averaging the $\pm$30 degrees on either side of the angle of the peak of the asymmetry if present, or around the major axis position angle for axisymmetric disks. The radial profiles provide the radial locations of the dust rings, in combination with more detailed analysis from the literature (after correction for the new \emph{Gaia} distances). These radii are also listed in Table \ref{tbl:sample} under the $R_{\rm dust}$ column. At each radial location, the azimuthal profile is extracted using a radial width of half the beam size. Figure \ref{fig:azimprofiles} presents both the radial and azimuthal profiles after normalization, where the latter are split in asymmetric and non-asymmetric structures. In the asymmetric curves the profiles are normalized to the flux at the opposite side of the asymmetric peak: in case of a non-detection on that side, a 3$\sigma$ upper limit is assumed for the normalization. 
Note that disks with multiple rings appear multiple times, with one curve for each ring. In the radial plots, the deprojected beam profile is overplot at the location of the dust ring, to show how well the ring is resolved radially. 

Contrasts in the asymmetric rings between peak and opposite side range from $\sim$3 to 395. 
The disks of SR~21 and CQ~Tau contain two asymmetric features along the same ring. The disk of SR~21 is not well resolved radially and higher resolution ALMA data show that these asymmetries are in fact more pronounced \citep[][Muto et al. in prep.]{Muro-Arena2020}. Also for the CQ~Tau disk the asymmetries are moderate, and higher resolution images (Benisty et al., in prep.) confirm these asymmetric features to be real.
Dust continuum asymmetries are marginally optically thick ($\tau\sim$0.5, using the brightness temperature and the expected temperature at the dust ring location, which is computed using Eqn 4 in \citet{Francis2020}), so the intensity contrasts do not directly correspond to density contrasts.
The symmetric rings show moderate azimuthal variations with values between 1 and 2 which may be optical depth effects in combination with the orientation of the disks or minor asymmetric dust traps.

Both MWC~758 and AB~Aur show a clear indication of an eccentric dust cavity, as the inner dust disk detected in the ALMA continuum image is offset from the center of the disk.

The azimuthal contrast in the images cannot be compared reliably across the different images, as its value depends on the beam size and SNR along the ring. Instead, the asymmetries can be described reasonably well by 2D Gaussian profiles in the radial and azimuthal direction, where the full-width-half-maximum (FWHM) is a measure of the asymmetric nature of the ring. Each asymmetric disk is fit in the visibility plane with a parametrized model $I(r,\phi)$ including such Gaussians. The visibility curves, parametrization and best fits are given in the Appendix \ref{sct:viscurves}. For HD135344B, a detailed visibility analysis was already performed by \citet{Cazzoletti2018}. Using the $\sigma_{\phi}$ values, we compute the FWHM of each asymmetric feature. For the axisymmetric disks, the FWHM is set to 360$^{\circ}$. The FWHM is listed in Table \ref{tbl:sample}.

\subsection{Gas structure}
\label{sct:gas}

\begin{table*}[!ht]
\caption{Gas gap properties}
\label{tbl:cogaps}
\begin{tabular}{lllllllllll}
\hline
Target	&	ALMA program	&	Line	&	Beam size	&	$R_{peak}$	&	$R_{CO}$	&	$R_{gap}$	&	$R_{gascav}$	&	$\Sigma_{acc}$ (1 au)	&	Ref CO analysis	\\
&&&(")&(au)&(au)&(au)&(au)&($\times10^3$ g cm$^{-2}$)&\\
&&&&(1)&(2)&(3)&(4)&(5)&(6)\\
\hline
IRS48	&	2013.1.00100.S	&	$^{13}$CO 6-5	&	0.19x0.15	&	40	&	29	&	22	&	31	&	0.80	&	1	\\
HD142527	&	2011.0.00318.S	&	$^{13}$CO 3-2	&	0.61x0.48	&	118	&	84	&	65	&	50\footnote{derived from threshold detectability of $^{12}$CO at 10$^{-2}$ g cm$^{-2}$.}		&	7.70	&	2	\\
ABAur	&	2012.1.00303.S	&	$^{13}$CO 3-2	&	0.37x0.23	&	117	&	84	&	64	&	98	&	26.00	&	3	\\
MWC758	&	2012.1.00725.S	&	$^{13}$CO 3-2	&	0.22x0.19	&	40	&	29	&	22	&	20$^a$	&	9.10	&	4	\\
HD135344B	&	2012.1.00158.S	&	$^{13}$CO 3-2	&	0.26x0.21	&	27	&	19	&	15	&	28	&	9.50	&	1	\\
SR21	&	2012.1.00158.S	&	$^{13}$CO 3-2	&	0.23x0.19	&	-	&	-	&	7\footnote{derived from rovibrational CO line \citep{Pontoppidan2008}}&	7	&	2.90	&	1	\\
CQTau	&	2017.1.01404.S	&	$^{13}$CO 2-1	&	0.15x0.15	&	23	&	16	&	13	&	20	&	1.00	&	5	\\
DoAr44	&	2012.1.00158.S	&	$^{13}$CO 3-2	&	0.31x0.29	&	26	&	19	&	14	&	24	&	1.90	&	1	\\
J1604-2130	&	2013.1.01020.S	&	$^{13}$CO 2-1	&	0.28x0.24	&	68	&	49	&	37	&	35$^a$	&	0.10	&	6	\\
LkCa15	&	2012.1.00870.S	&	$^{13}$CO 3-2	&	0.28x0.21	&	45	&	32	&	25	&	35	&	1.20	&	7	\\
PDS70	&	2017.A.00006.S	&	$^{12}$CO 3-2	&	0.08x0.06	&	41	&	29	&	23	&	22\footnote{derived from resolved $^{12}$CO profile \citep{Keppler2019}}	&	0.04	&	8	\\
Sz91	&	2013.1.01020.S	&	$^{13}$CO 2-1	&	0.25x0.22	&	52	&	37	&	29	&	32	&	0.58	&	9	\\
HD169142	&	2013.1.00592.S	&	$^{13}$CO 2-1	&	0.37x0.22	&	-	&	-	&	12, 42\footnote{estimated from dust ring locations} 	&	60	&	2.90	&	10	\\
DMTau	&	2017.1.01460.S	&	$^{12}$CO 2-1	&	0.10x0.10	&	-	&	-	&	12$^d$	&	-	&	1.40	&	11	\\
HD163296	&	2013.1.00601.S	&	$^{13}$CO 2-1	&	0.27x0.19	&	-	&	-	&	-	&	-	&	7.60	&	12	\\
TWHya	&	2012.1.00422.S	&	$^{13}$CO 3-2	&	0.54x0.35	&	-	&	-	&	-	&	-	&	0.32	&	13	\\

\hline
\end{tabular}
Explanation of columns. (1) Radial peak of $^{13}$CO emission (or $^{12}$CO when $^{13}$CO is not available) at the outer edge of the gas gap; (2) Gas gap edge, as derived from $R_{\rm peak}$ and the relations in \citet{Facchini2017gaps}; (3) Gap gap minimum, as derived from $R_{\rm peak}$ and the relations in \citet{Facchini2017gaps}; (4) Gas cavity edge from parametrized gas surface density model from the literature, corrected for the \textit{Gaia DR2} distance; (5) Gas surface density at 1 au, using the accretion rate and Equation \ref{eqn:macc}; (6) Reference of the analysis of the CO isotopologues from the literature: 
1) \citet{vanderMarel2016-isot}, 2) \citet{Boehler2017}, 3) \citet{Pietu2005}, 4) \citet{Boehler2018}, 5) \citet{Ubeira2019}, 6) \citet{Dong2017}, 7) van der Marel et al. in prep., 8) \citet{Muley2019}, 8) , 9) \citet{vanderMarel2018a}, 10) \citet{Fedele2017}, 11) Francis et al. in prep., 12) \citet{vanderMarel2018-hd16}, 13) \citet{Kama2016twhya} 
\end{table*}

CO isotopologue images of transition disks reveal that the gas cavity radii are well within the dust cavity radii. Gas surface density profiles have been derived from these CO images, in order to quantify the depth and width of these gas gaps that can be used to derive information about possible embedded companions. Unlike the dust ring, which generally has a sharp inner edge due to the trapping \citep{Pinilla2018tds},  the gas gap edge is not expected to be sharp, but has been shown to have a gradual drop in density, consistent with clearing by a companion, with the minimum at the location of the companion and the dust trapped at the outer edge. We have defined $R_{\rm dust}$ as a maximum in the intensity profile of dust emission (see previous Section), and the location of the gas gap $R_{\rm gap}$ as the minimum in the gas surface density profile. 

\begin{figure}[!ht]
\includegraphics[width=\textwidth]{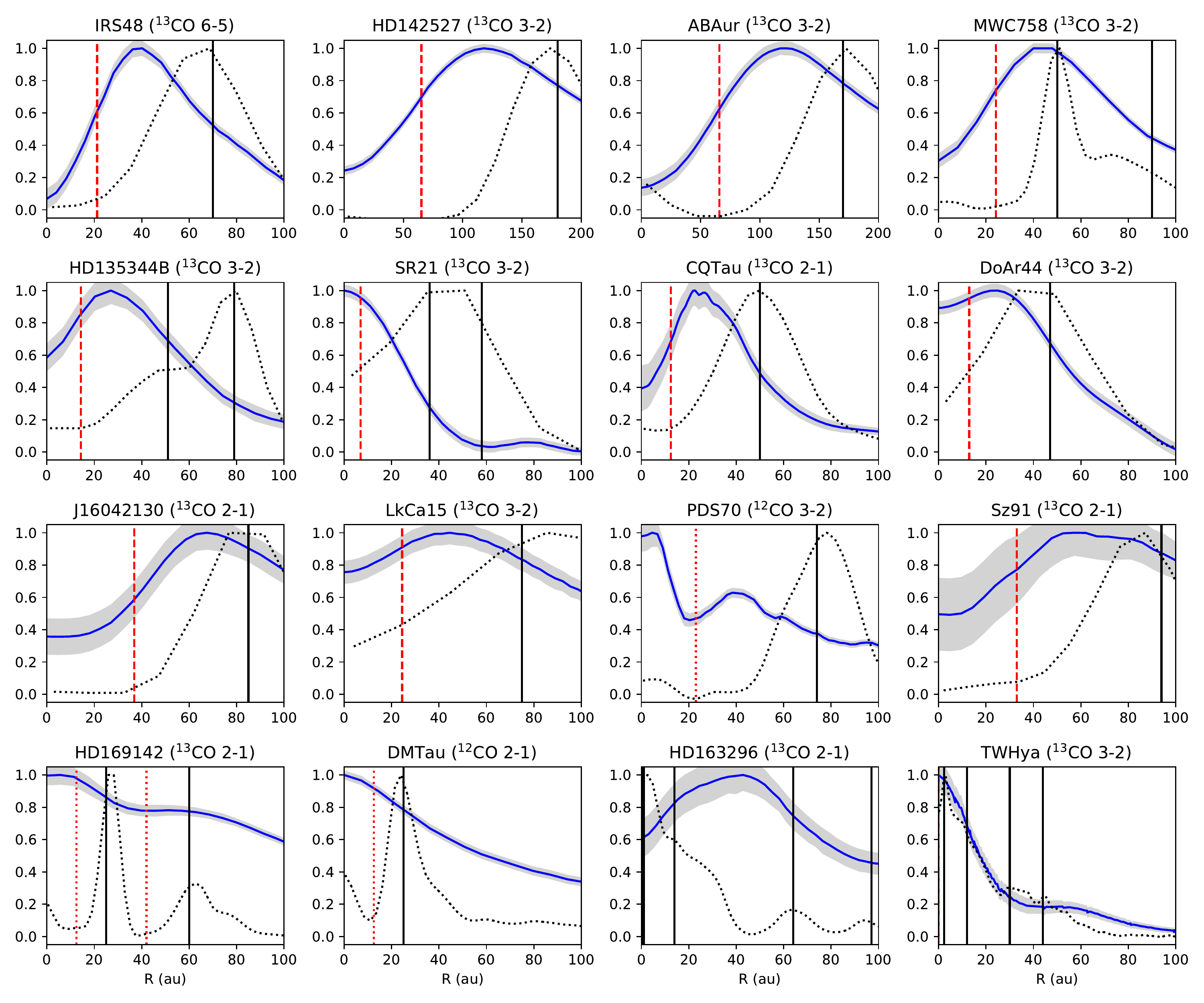}
\caption{Normalized intensity profiles of $^{13}$CO profiles of each target. The title indicates which line is used. Black dotted profiles show the radial dust profiles and black vertical lines indicate the locations of the dust ring(s). The grey area indicates the noise level. Red dashed lines indicate the derived gap radius $R_{gap}$ (see text). Dotted red lines are used for targets where this analysis could not be done: for SR~21 the gap radius is taken from the analysis of the rovibrational CO line \citep{Pontoppidan2008}; for PDS~70 the gap radius is taken from the spatially resolved $^{12}$CO profile \citep{Muley2019}; for the last four targets no information could be derived from the CO intensity profiles on the gap radius and they are estimated to be located in between the dust rings.}
\label{fig:COgaps}
\end{figure}

In model fitting of CO images, the gap in the surface density profile is usually parameterized \citep[e.g.][]{vanderMarel2016-isot}, in order to limit the number of free parameters. Unfortunately the parametrization is not the same in different studies and a comparison across the sample is challenging. Furthermore, the parametrization in early studies usually contained unphysical sharp edges at the gas cavity radii. Therefore, we re-evaluate the gas surface density profiles and the gap radii $R_{\rm gap}$ by analyzing the normalized azimuthal averaged intensity profiles of $^{13}$CO for each target (Figure \ref{fig:COimages} and \ref{fig:COgaps}). The properties and origin of each CO image are summarized in Table \ref{tbl:cogaps}. 

None of the $^{13}$CO intensity profiles reveal a gap structure as expected from simple planet-disk interaction models: the minimum is located at the center of the disk. However, the spatial resolution of these images is limited (usually on the order of the size of the gap, $\sim$0.2-0.3" or $\sim$30-50 au), so the inner gas disk emission, if detected, is either unresolved \citep[see also Figure 9 in][]{vanderMarel2018a} or the emission is lower than predicted by these models due to either a decrease in temperature, insufficient knowledge of processes inside the planet orbit leading to a further depletion of the gas, or both. 
Massive planets on eccentric orbits will significantly deplete the inner gas disk compared to regular planet-disk interaction models where planets are held fixed on circular orbits \citep{Muley2019}.

The deep high resolution $^{12}$CO images of PDS70 do reveal a clear gas gap \citep{Keppler2019} and high resolution dust continuum images reveal that inner dust disk are common in transition disks \citep{Francis2020}, suggesting that in fact many of these disks indeed harbor gaps rather than cavities. Also the high accretion rates in transition disks (comparable to those of full disks) suggest a higher gas surface density close to the star \citep{Manara2014,Francis2020}. \citet{Bosman2019} find evidence that the CO temperature in Herbig disks must be significantly lower than physical-chemical models predict to explain the ratios between different rovibrational lines. Such a decrease in temperature may also be a reason that $^{13}$CO remains undetectable in the inner parts of the disk and thus we assume in this study that all disks in fact harbor gas gaps. 

In this work, we derive the gap location directly from the $^{13}$CO profile across the sample. Since $R_{\rm gap}$ cannot be directly constrained for all sources, we rely on two additional quantities, $R_{\rm CO}$ and $R_{\rm peak}$ to infer it. In model simulations, the gap edge $R_{\rm CO}$ is defined \citep{Rosotti2016,Facchini2017gaps} as the radius $R$ where the normalized intensity $\bar{I}(R)< 1-0.66(1-\bar{I}_{min})$. The low spatial resolution of our $^{13}$CO observations does not allow to measure this parameter directly, as the gap remains unresolved and it is unclear whether potential inner disk gas emission is confused or is highly decreased. As an alternative, we measure the location of $R_{\rm peak}$, the peak in the integrated $^{13}$CO emission. Inspection of the results in \citet{Facchini2017gaps} shows that $R_{\rm peak}$ is approximately 1.4 times larger than $R_{\rm CO}$. Figure 11 and 12 in \citet{Facchini2017gaps} provide the relations between $R_{\rm CO}$ and $R_{\rm gap}$, and for planet masses 5-15 $M_{\rm Jup}$ the ratio $R_{\rm CO}/R_{\rm gap}$ = 1.3. Using these relations, we derive the location of $R_{\rm peak}$, $R_{\rm CO}$ and $R_{\rm gap}$, as listed in Table \ref{tbl:cogaps}. The derived gap radii are well inside the dust cavity at typically 10-20 au radius. If the relations between $R_{peak}$ and $R_{gap}$ are invalid because of eccentricity, $R_{gap}$ is likely even further in. Unfortunately no grid of models exists for planet-disk interaction models including eccentricity with predictions for the CO emission, so we have to make the assumption here that the radial gap shape remains similar. For SR~21, no gap was resolved in $^{13}$CO and the inner radius derived from the rovibrational emission \citep{Pontoppidan2008} is assumed. For PDS~70, no $^{13}$CO data are available but the $^{12}$CO profile directly reveals the gas gap in the image \citep{Keppler2019}. We notice that the ratios above still recover the gap radius at almost the exact same location for the $R_{\rm peak}$ of $^{12}$CO in this case. For HD~169142, DM~Tau, HD~163296 and TW~Hya no gas gaps were resolved and the gap locations are estimated to be located in between the dust ring radii. The drop in emission in the center of HD~163296 is caused by continuum oversubtraction \citep{Isella2016}.

Second, we present the first moment maps (velocity maps) of $^{12}$CO data in Appendix \ref{sct:kinematics} for each of our targets. A twist pattern (deviation from Keplerian rotation) in the inner part of the disk points towards a misalignment between inner and outer disk, or a warp. Such a misalignment can be explained by the presence of a massive companion ($>1 M_{\rm Jup}$), which breaks the disk, leading to a different precession of inner and outer disk \citep[e.g.][]{Facchini2017warps,Zhu2019}. An even stronger misalignment can be induced as a result of a secular resonance between the companion and the disk \citep{OwenLai2017}. Also radial flows of gaseous material from the outer to the inner disk have been proposed to explain the twist pattern \citep[e.g.][]{Price2018}, but it is almost impossible to distinguish observationally from a misalignment \citep{Rosenfeld2014} and not unique for substellar companions such as found in HD~142527, as lower-mass companions can result in radial flows as well \citep{Calcino2020}. Misalignment has been discovered independently in several targets through shadows \citep[e.g.][]{Marino2015}, dippers \citep[e.g.][]{Ansdell2016b} and direct measurements of the inner dust disk orientation \citep[e.g.][]{Francis2020}.

For the targets in this study, a warp was confirmed for 4 targets, that were previously found in the literature:  IRS~48, HD~142527, MWC~758 and J1604-2130. AB~Aur appears to show non-Keplerian motion as well on larger scales, but this is most likely due to the strong contributions from the spiral arms detected in $^{12}$CO \citep{Tang2017}. For the other disks no warp was detected, but we present a comprehensive overview of the properties of the observations suggesting that warps are impossible to detect with the available spatial/spectral resolution (Table \ref{tbl:kinematics}). The Table also provides references for other studies suggesting misalignment based on other data. Overall, all targets possibly have a misalignment between the inner and outer disk, pointing towards the presence of a massive companion. Derivation of the mass of the companion requires detailed knowledge of the viscosity, scale height and precession time \citep[Figure 12 in][]{Zhu2019}, so no quantitative information can be derived from these maps.

\subsection{Companions}
\label{sct:comps}

\begin{figure}[!ht]
\includegraphics[width=0.9\textwidth]{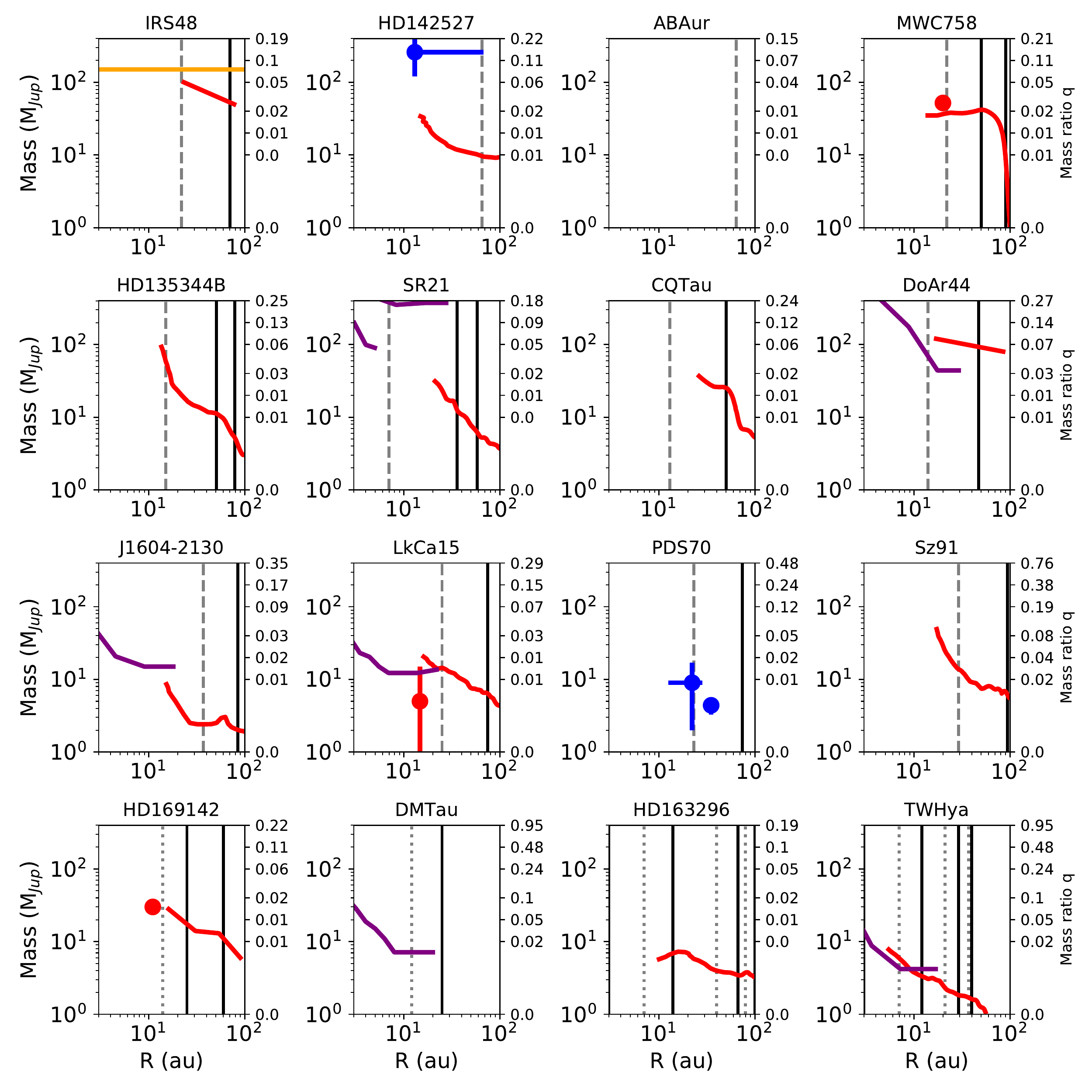}
\caption{Companion candidates and limits for each target (see Table \ref{tbl:sample} for references). Red curves show the results from coronagraph studies, purple lines from Sparse Aperture Masking studies, and orange for lunar occultation (see text). Blue symbols show confirmed companions, whereas red symbols are unconfirmed candidates. The black solid lines indicate the locations of the dust rings, whereas the dashed gray lines indicate the gas gap locations as derived from the $^{13}$CO in Section \ref{sct:gas}. Dotted gray lines indicate estimates of the gas gaps in between the dust rings.}
\label{fig:companions}
\end{figure}

Many transition disks have been studied in direct imaging searches for embedded companions. Figure \ref{fig:companions} presents the best known limits for companions in the disks in our sample, in comparison with the locations of the dust rings and gas gaps. The right y-axis provides the mass ratio with respect to the stellar mass in percentage. 
References for high contrast direct imaging searches are provided in Table \ref{tbl:sample}, additional data are discussed below. The mass upper limit curves are computed in these works by comparing the brightness limit with evolutionary models such as BT-SETTL \citep{Allard2014} or COND \citep{Baraffe2003} assuming the age of the system and hot-start models. When values for both models were computed, we adopt the BT-SETTL results. Companion detections are marked with red (unconfirmed) and blue (confirmed) circles.  The evolutionary models for upper limits and most candidates do not include contributions from a circumplanetary disk (CPD), which could dominate the brightness and can lower the mass limit estimates by a factor of 10 \citep{Zhu2015Nature}. Also extinction, age estimate and choice of evolutionary model may affect the derived mass estimates and limits. 

For PDS~70 and LkCa15, H$\alpha$ and multi-wavelength data (SED) provide constraints on the CPD, which means that their estimated companion mass is much lower than the typical contrasts in other disks. Although the SED of HD~142527B might be explained with a planetary companion with a CPD as well \citep{Brittain2020}, the companion mass is very likely substellar based on a proper motion study of the primary star \citep{Claudi2019} and on previous fits of the SED and SINFONI H+K spectrum to BT-SETTL models \citep{Lacour2016,Christiaens2018}.

For IRS~48, upper limits on companion brightness were measured by \citet{Ratzka2005} with speckle imaging at 0.15" and 0.5" (20 and 67 au), converted to mass limits of 100 and 50 $M_{\rm Jup}$ by \citet{Wright2015}, which are interpolated in our contrast curve. In addition, \citet{Simon1995} measured a K-band contrast in the regime 0.02-1" ($\sim$3-135 au) using the lunar occultation method, which was converted to a mass limit of 150 $M_{\rm Jup}$ \citep{Wright2015}. The latter is marked with an orange curve in Figure \ref{fig:companions}. Also for DoAr44 we add limits from \citet{Ratzka2005} converted by \citet{Wright2015}. For AB~Aur no contrast curves have been derived to our knowledge.  
For SR~21, \citet{Sallum2019} performed a detailed analysis using sparse aperture masking detecting features around 7 au, but modeling showed that these were more consistent with an inner dust ring rather than a companion. The features require a warped inner disk or spiral features. 
For LkCa~15, the companion candidates c and d identified by \citet{Sallum2015} have been suggested to originate from scattered light by inner disk material in follow up studies \citep{Thalmann2016,Currie2019}, but we include LkCa~15b as a companion candidate due to its detection at H$\alpha$ and the lack of polarized emission at its location.

In additions to the limits presented in Figure \ref{fig:companions}, we show additional constraints from a brown dwarf survey through sparse aperture masking for SR~21, DoAr44, J1604-2130, LkCa15, DM~Tau and TW~Hya \citep{Kraus2008, Kraus2011, Evans2012, Cheetham2015} for the inner 0.15" of the disk and additional limits for SR~21 for the inner 5 au (Sallum, private communication) using the data from \citet{Sallum2019}. Unlike coronagraphy, sparse aperture masking allows the detection of companions at angular separations well within the diffraction limit down to a few au at typical disk distances \citep[e.g.][]{Sallum2019spie}. Mass limits were derived using a range of evolutionary models using the procedure described in \citet{KrausHillenbrand2007}.

A handful of indirect estimates of companion candidates are known from the literature, using a range of techniques. 
\begin{itemize}
\item \citet{Baines2006} claim that AB~Aur has an accreting binary companion at 82-489 au separation using H$\alpha$ spectro-astrometry, which could be either inside or outside the cavity and with unknown mass. 
\item \citet{Boccaletti2020} deduce a planet of 4-13 $M_{\rm Jup}$ at 30 au in AB~Aur based on a spiral arm twist. 
\item \citet{Gratton2019} find a tentative detection of  a 3 $M_{\rm Jup}$ companion at 38 au separation in HD~169142.
\item \citet{Willson2016} find a tentative detection of a companion at 6 au separation in DM~Tau using sparse aperture masking without significant sky rotation.
\item \citet{Pinte2018} and \citet{Teague2018} find evidence for $\sim$2 $M_{\rm Jup}$ planets at 83, 137 and 260 au in HD~163296 using deviations of Keplerian motion in $^{12}$CO channel maps.
\item \citet{Calcino2019} and \citet{Poblete2020} claim substellar companions with mass ratios of $\sim$0.2 at 10 and 30 au in IRS~48 and AB~Aur, respectively, to explain the kinematics in the system.
\end{itemize}
Due to their more speculative nature these candidates are not marked in the contrast curves in Figure \ref{fig:companions}. Follow-up observations are required to confirm their existence.

\begin{figure}[!ht]
\includegraphics[width=\textwidth]{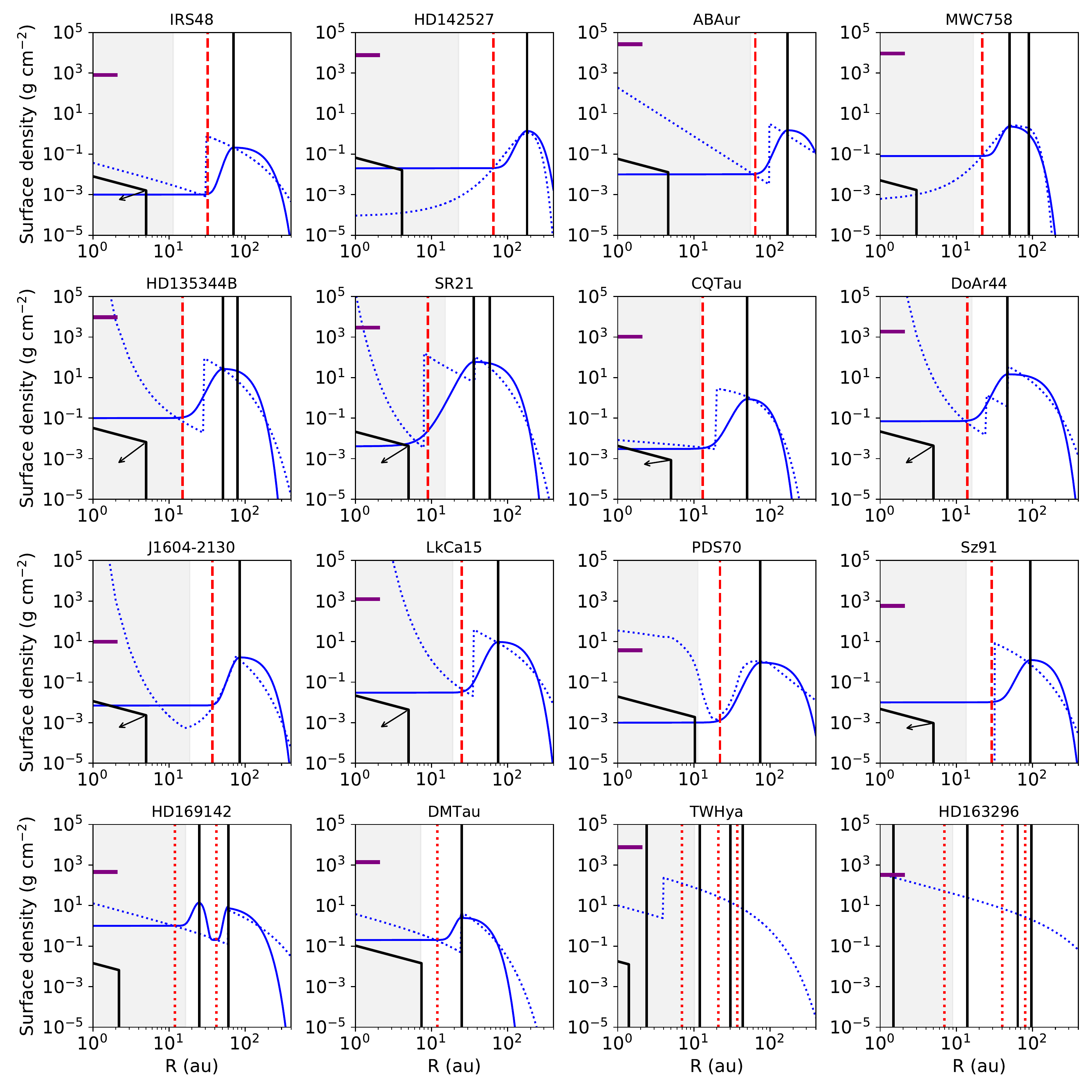}
\caption{Estimated gas surface density profiles from our $^{13}$CO analysis (solid blue). The dotted blue profiles show the best fit surface density profiles from the literature from a full analysis of the CO isotopologues (references in Table \ref{tbl:cogaps}). Black solid lines indicate the dust ring locations and the inner disk dust profiles derived by \citet{Francis2020}. The red dashed lines indicate the location of the derived gap. Dotted red lines indicate estimates of the gas gap radii in between the dust rings.  A purple marker is set at the expected gas surface density at 1 au considering the mass accretion rate of the star (see text). The gray area indicates the regime where the ALMA CO data remains unresolved (beam radius) and the gas surface density profile thus remains highly uncertain.}
\label{fig:gaussprofiles}
\end{figure} 

\clearpage
\newpage

\section{Analysis}
\label{sct:analysis}

\subsection{Gas gaps}
\label{sct:gasprofiles}

In Figure \ref{fig:gaussprofiles} we present the estimated gas surface density profile, described as a Gaussian centered on $R_{\rm dust}$ and an inner width consistent with the $R_{\rm gap}$ location. $R_{dust}$ is chosen as the outer edge of the gas gap as the pressure maximum is thought to be located there. $R_{peak}$ is inwards of $R_{dust}$ but this can be understood by the radial temperature dependence: whereas the gas surface density is decreasing, the $^{13}$CO emission remains partially optically thick and peaks further inwards. For DM~Tau and SR~21, no $R_{\rm peak}$ could be measured from the CO profile and it was estimated to be located at $\sim$75\% of the dust ring radius. The profile is scaled to the derived gas surface density profile from the literature based on combined $^{13}$CO and C$^{18}$O data (see references in the last column in Table \ref{tbl:cogaps}) to match the surface density at peak, gap and outer disk locations. The gap radii in the literature profiles were rescaled to the \emph{Gaia} distances. The literature surface density profile is overplotted as blue-dotted line and Table \ref{tbl:cogaps} lists the gas cavity radii $R_{\rm gascav}$ from these profiles from the literature. For PDS~70 and J1604.3-2130 the CO data were fit with a gap-like profile (material inside the companion orbit). 
For HD~142527 and MWC~758 the gas surface density profile was described as a Gaussian in the literature analysis \citep{Boehler2017,Boehler2018}: $R_{gascav}$ is taken as the radius where the density drops below 10$^{-2}$ g cm$^{-2}$ where $^{12}$CO becomes optically thin. For AB~Aur no detailed physical-chemical model was used and the CO abundance was taken to be constant throughout the disk, so the derived profile remains highly uncertain. TW~Hya and HD~163296 do not have a resolved \emph{inner} gas gap, so no values are provided here. HD~163296 does show gas gaps in the outer disk \citep{Isella2016},  but the depth and width remain highly uncertain and we refrain from including them in the plot \citep{vanderMarel2018-hd16}.

The inner part of the gas disk remains largely unconstrained by our analysis of the CO images due to spatial resolution. We indicate the unconstrained regime with a gray area between 0 and the beam radius in Figure \ref{fig:gaussprofiles}. This representation reveals that for none of the disks, except PDS~70, the distinction between a gas gap or gas cavity can be made. However, inner dust disks have been detected in about half of the sample \citep{Francis2020} and all disks show signs of significant gas accretion onto the star, which makes it most likely that gas is still present in the inner disk as well and the gas `cavities' are in fact gas gaps, consistent with clearing by a companion. The derived inner disk dust profiles from \citet{Francis2020} and the expected gas surface density based on the accretion are included in the plot to reflect this. The expected gas surface density close to the star can be estimated indirectly assuming a viscous disk model from the stellar accretion rate by \citet{Manara2014}:
\begin{equation}
\label{eqn:macc}
\Sigma_g(r)= \frac{\dot{M} 2 m_p}{3\pi\alpha k_BT(r)}\sqrt{\frac{GM_*}{r^3}},
\end{equation}
with $\Sigma_g(r)$ the gas surface density, $\dot{M}$ the accretion rate, $m_p$ the proton mass, $\alpha$ the viscosity, $k_B$ the Boltzmann constant, $G$ the gravitational constant, $M_*$ the stellar mass and $T(r)$ the temperature profile, for which we use \citep[][]{Dullemond2001}:
\begin{equation}
T(r) = \Big(\frac{\phi L_*}{8\pi\sigma_Br^2}\Big)^{1/4} = \sqrt[4]{\frac{\phi L_*}{8\pi\sigma_B}} \frac{1}{\sqrt{r}}
\end{equation}
with $L_*$ the stellar luminosity, $\phi$ the flaring angle (taken as 0.02) and $\sigma_B$ the Stefan-Boltzmann constant. The stellar properties are taken from \citet{Francis2020}. We compute the expected local gas surface density $\Sigma_{acc}$ at 1 au derived from the accretion rate assuming $\alpha=10^{-3}$ (see Table \ref{tbl:cogaps}), and overplot this value on the derived profiles (Figure \ref{fig:gaussprofiles}).

\begin{figure}[!ht]
\centering
\includegraphics[width=0.5\textwidth]{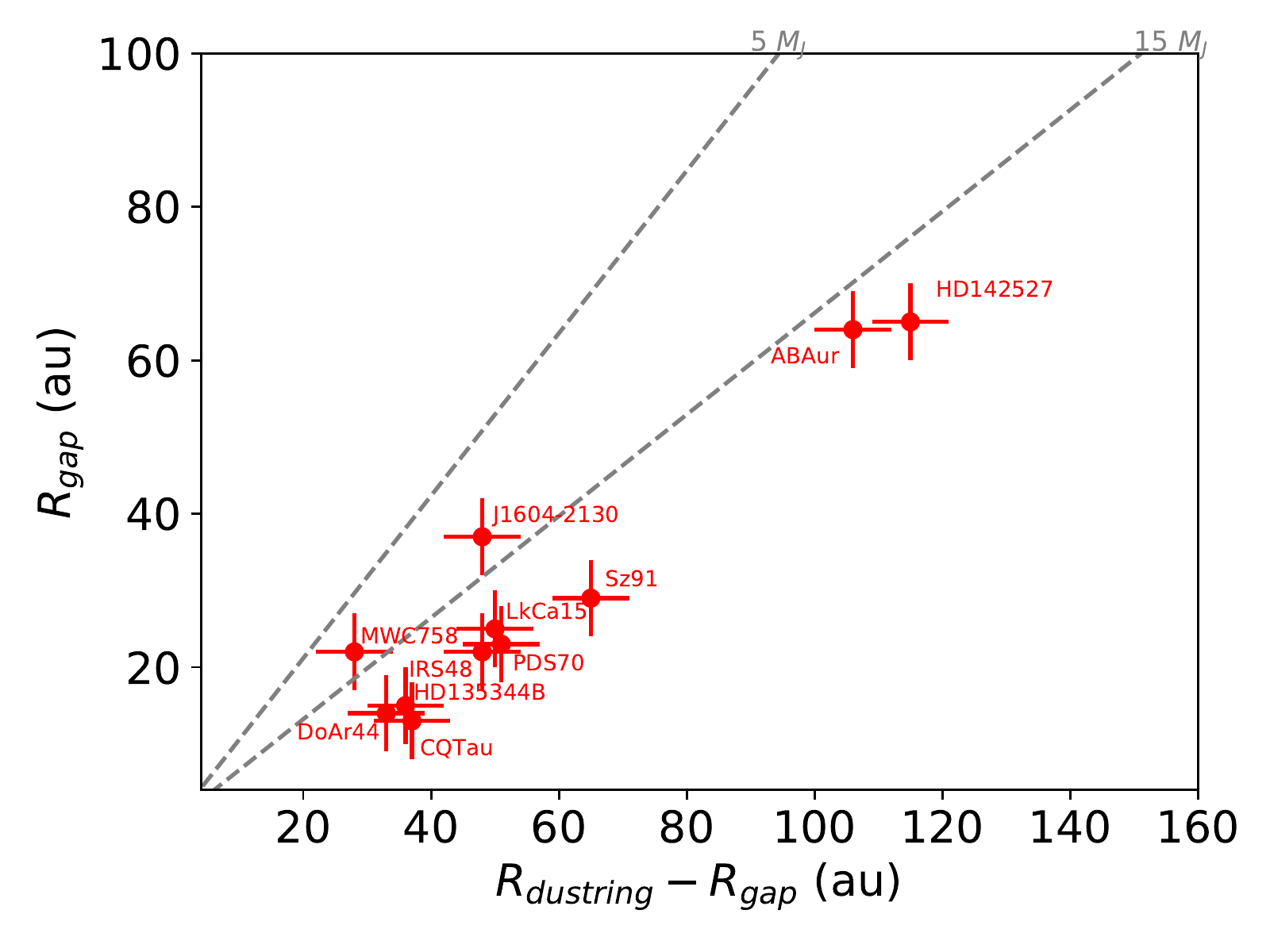}
\caption{Gas gap vs dust ring radius for each target, based on analysis of CO observations (see Table \ref{tbl:sample} for references). The gray dashed lines indicate the regime where the gas gap is expected to be caused by a companion of 5 and 15 $M_{\rm Jup}$ on a circular orbit, following the relations derived by \citet{Facchini2017gaps}. Most disks fall below the 15 $M_{\rm Jup}$ line in this case and would thus have companions above that threshold, in the (sub)stellar regime}. \label{fig:gapratio}
\end{figure}

In order to estimate the companion mass assuming a single planet, the millimeter dust radius $R_{\rm dust}$ is compared with the derived gas gap radius $R_{\rm gap}$.  Figure \ref{fig:gapratio} presents this relation. Best fit relations between the ratio and the planet mass were derived by \citet{Facchini2017gaps} for planet masses in between 1  and 15 $M_{\rm Jup}$ for an average between $\alpha=10^{-3}$ and $\alpha=10^{-4}$, but as no models were run for higher mass companions, this relation cannot be used to estimate accurate masses for these ratios. Also, Facchini et al. do not consider eccentric orbits which significantly increase the separation between the dust ring and the gas gap \citep{Muley2019}. The majority of our disks lie in the regime $>15 M_{\rm Jup}$, suggesting that they contain planets above this threshold, in the brown dwarf regime. The ratio between $R_{\rm dust}-R_{\rm gap}$ and $R_{\rm gap}$ for our sources is typically between 1.3 and 2.5. Outliers are CQ~Tau and SR~21 with even higher ratios (3-4), suggesting very massive companions, potentially (sub)stellar. 

A similar comparison was made between $R_{\rm dust}$ and the inner edge of the scattered light gap for a sample of transition disks \citep{Villenave2019} using the planet mass relations derived by \citet{deJuanOvelar2013} for $\alpha=10^{-3}$. For the overlapping targets, their estimates of companion mass are consistent with ours, with several substellar mass companions. Exceptions are IRS~48 and Sz~91, for which they use non-scattered light observations for which the planet relation does not hold, and LkCa~15 for which the inner edge is more challenging to determine in scattered light due to its high inclination. Also several other disks in their sample (not in our work) that are claimed to be in the planetary regime might suffer from high inclination.

\subsection{Stokes numbers}
\label{sct:stokes}

\begin{figure}[!ht]
\centering
\includegraphics[width=0.45\textwidth,trim=0 0 0 0]{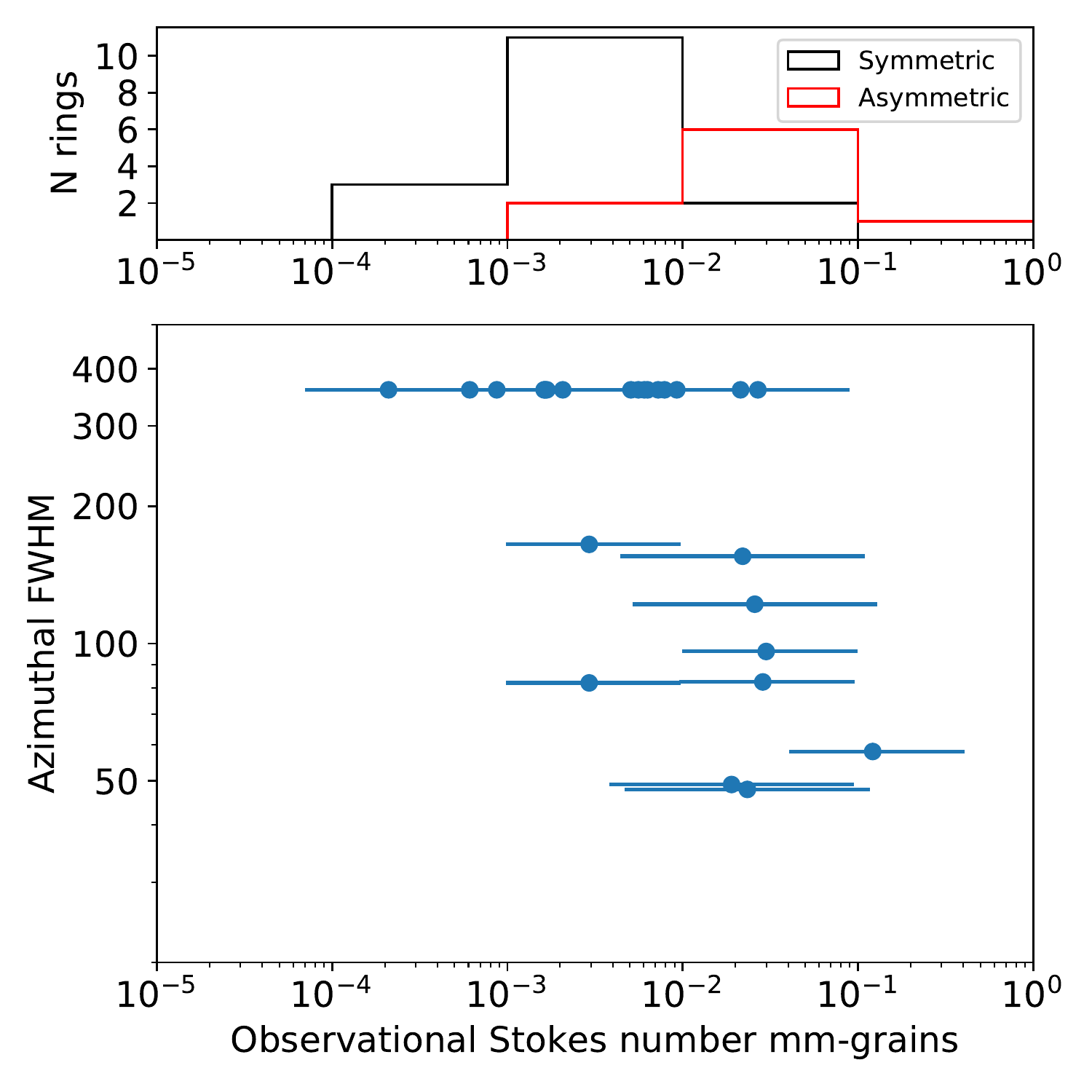}
\includegraphics[width=0.45\textwidth,trim=0 0 0 0]{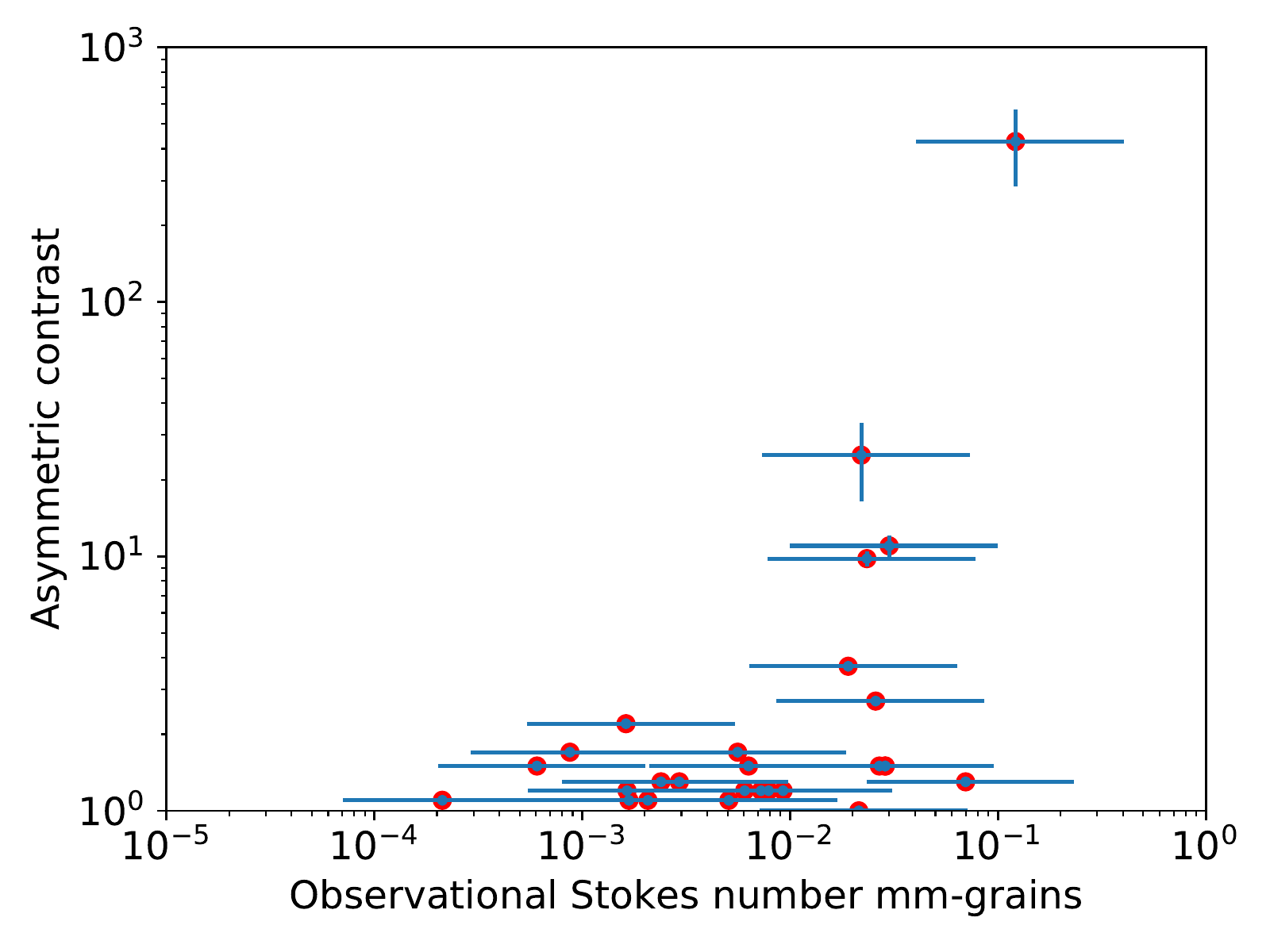}
\caption{{\bf Left:} Azimuthal extent as function of Stokes number of millimeter grains at location dust ring/asymmetry for each of our targets, computed from the visibility analysis. The Stokes number has been computed using the observing wavelength and the local gas surface density (see text for details). The top panel shows the distribution of the targets. {\bf Right:} Azimuthal contrast as function of Stokes number, computed from the images.}
\label{fig:stokesdata}
\end{figure}

With the azimuthal profiles fit in Appendix \ref{sct:viscurves} and the gas surface density profiles described in Section \ref{sct:gasprofiles} we construct a plot of the azimuthal FWHM as a function of Stokes number of the traced dust grains, for the radial location of the dust. The Stokes number $St$ as defined in Eqn \ref{eqn:stokes} cannot be used directly, as dust continuum emission is originating from a large range of grain sizes (and Stokes numbers). Therefore, we use a simplification with the assumption that particles with size $a_{\rm grain}=\lambda_{obs}/2\pi$ \citep{Draine2006} are the primary contributor at observing wavelength $\lambda_{obs}$, and introduce the observational Stokes number:
\begin{equation}
\label{eqn:stobs}
St_{obs} = \frac{\lambda_{\rm obs}\rho_s}{4\Sigma_{\rm gas}(R_{\rm dust})}
\end{equation}
with the gas surface density $\Sigma_{\rm gas}$ at the location of the dust ring $R_{\rm dust}$. The result is shown in Figure \ref{fig:stokesdata}. For the uncertainties we assume an uncertainty of a factor 3 on $\Sigma_{\rm gas}(r)$, based on the typical uncertainty on the gas surface density based on CO isotopologue data as derived by \citet{Woitke2019}. Although other grain sizes than $a_{\rm grain}$ may contribute to the emission which might add additional uncertainty, this is not considered an issue as all Stokes numbers are computed in the same way. As it is reasonable that the grain size distributions are similar across the sample (under the assumptions that the disks have similar ages and are evolving in similar physical environments), it would thus shift all data points in the same direction and the trend would remain the same.

Figure \ref{fig:stokesdata} shows that axisymmetric disks have low values of the observational Stokes number, but asymmetric features are all located at $St_{obs}>10^{-2}$. The derivation of gas surface density from CO isotopologue data remains uncertain, in particular due to problems with our knowledge of the carbon budget in disks \citep{Kama2016,Miotello2019}. In addition we show the dependence of the azimuthal contrast from the images on the observational Stokes number which also shows a distinction between symmetric and asymmetric disks. As this trend might be affected by imaging artefacts it is not further discussed.

\begin{figure}[!ht]
\centering
\includegraphics[width=\textwidth,trim=100 0 100 0]{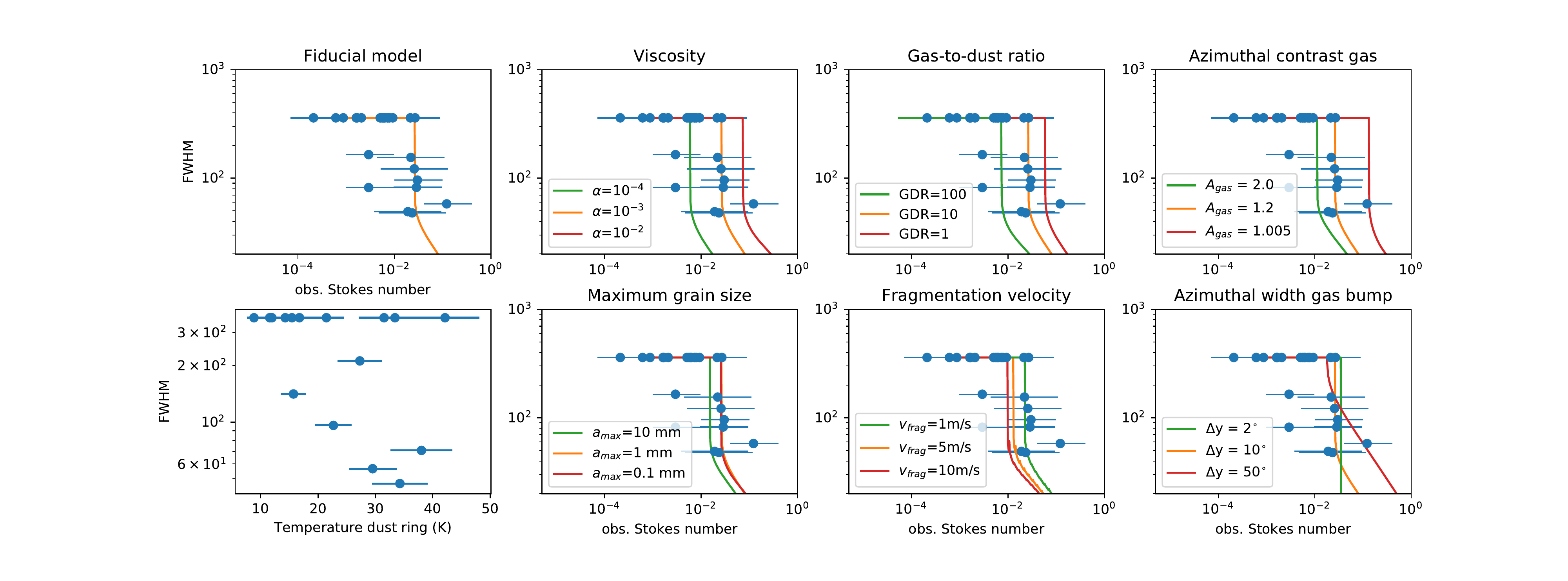}
\caption{Azimuthal extent as function of observational Stokes number of millimeter grains at location dust ring/asymmetry. The colored lines indicates the width expected from analytical relations of azimuthal dust trapping \citep{Birnstiel2013} with $\alpha=10^{-3}$ (see text). The orange line is considered as most consistent with the data. The plot in the lower left shows the relation between the width and the temperature at the location of the dust ring.}
\label{fig:stokes}
\end{figure}

To explore the physical implications of these results, we employ a model presented in \citet{Birnstiel2013} that analytically solves for the equilibrium of azimuthal drift and azimuthal mixing of dust particles. The azimuthal contrast of this model was shown to be in good agreement with 2D calculations of \citet{LyraLin2013}. Our model consists of three steps: 1) constructing a particle size distribution, 2) calculation of the azimuthal equilibrium density distribution of each particle size according to \citet{Birnstiel2013} and 3) calculating the dust intensity profiles and thus the azimuthal dust intensity contrast. For the first part, the particle size distribution, we tried two different choices, first a truncated power-law size distribution with an MRN exponent \citep{Mathis1977} up to a maximum particle size $a_\mathrm{max}$ and secondly, the steady-state distributions of \citet{Birnstiel2011}. For the second part, the azimuthal density distribution, we employed Eq.~8 of \citet{Birnstiel2013} and parameterized the azimuthal gas density as a constant density plus a Gaussian overdensity,

\begin{equation}
\Sigma_g(r, y) = \bar\Sigma_g(r) \frac{1 + (A-1) e^{-\frac{y^2}{2 r^2 \sigma
   ^2}}}{1 + (A-1)\frac{ \sigma  }{2 \sqrt{2 \pi }} \,\text{Erf}\left(\frac{\sqrt{2} \pi }{\sigma
   }\right)}
\end{equation}

where $y$ is the azimuthal coordinate going from 0 to $2\pi\,r$ and $\sigma_y$ is the azimuthal extent of the bump. $\Sigma_\mathrm{g}(r, y)$ is normalized such that the azimuthal average gives $\Sigma_\mathrm{g}(r)$.

The emission profile was then calculated assuming absoption opacity and a face-on, vertically isothermal disk, such that the intensity becomes $I_\nu = B_\nu(T_\mathrm{dust}) \left(1 - e^{-\tau}\right)$, where $\tau = \sum_i \Sigma_i \, \kappa_{\mathrm{abs},i}$ is the optical depth. $\Sigma_i$ and $\kappa_{\mathrm{abs},i}$ are the surface density and the absorption opacity of grain size $i$. Optical depth is thus taking into account for the comparison with the data. 
The final azimuthal FWHM is computed directly for the intensity profile of the model. Although this is a rather simple approach, it is sufficient for the purpose of reproducing the trend of azimuthal extent w.r.t. observational Stokes number.

We explore a number of parameters, including $\alpha$, gas-to-dust ratio, gas contrast and gas azimuthal extent and find the best fit, accompanied by two values on either side to show the dependence of the curve on the parameter. The threshold of the observational Stokes number where disks become asymmetric ($\sim 10^{-2}$ for most of our targets) depends on the disk properties, and may vary somewhat from disk to disk. Furthermore we explore whether the FWHM is set by either fragmentation (different fragmentation velocities) or by a default grain size distribution with a maximum grain size. Fragmentation velocities in lab experiments range from 1-10 m/s \citep[e.g.][]{BlumWurm2008} and even higher outside the snowline \citep{Wada2009}, although the latter has been called into question by recent experiments \citep{Steinpilz2019}. The fiducial model (orange) shows a possible combination based on a manual fitting procedure: $\alpha=10^{-3}$, gas-to-dust ratio = 10, gas contrast = 1.2, $a_{\rm max}$=1 mm and azimuthal $\sigma_y$ of 10$^{\circ}$. 
In addition, we plot the azimuthal FWHM as function of the temperature (at the dust ring) in this Figure.

\subsection{Spiral arms}
\label{sct:spirals}
A final aspect that is relevant for this discussion is the presence of spiral arms and the link with asymmetries. Table \ref{tbl:sample} indicates which disks show spiral arms in scattered light, with the references provided in the last column. About half of the sample (all asymmetric disks) shows spiral arms. For DM~Tau no scattered light imaging data is available and the spiral nature of HD~169142 remains uncertain as the spiral arms are seen in total scattered light \citep{Gratton2019}, through angular differential imaging but not in polarized scattered light \citep{Pohl2017,Bertrang2018}. The former technique suffers, unlike polarized differential imaging, from possible biases deriving from the disk emission self-subtraction, and in particular if the disk is seen face-on.

\subsubsection{Link with asymmetries}
\citet{Garufi2018} noticed that all asymmetric disks in ALMA show spiral arms in scattered light, and the apparent origin of one of the spiral arms in HD135344B in the dust asymmetry \citep{vanderMarel2016-spirals} suggests that these phenomena are physically linked if a vortex triggers the spiral arm \citep{Lovelace2014}. However, recent simulations show that spiral arms triggered by a vortex are unlikely to be detectable in scattered light \citep{Huang2019} and the spirals must have a different origin such as a companion.

\begin{figure}[!ht]\
\centering
\includegraphics[width=\textwidth]{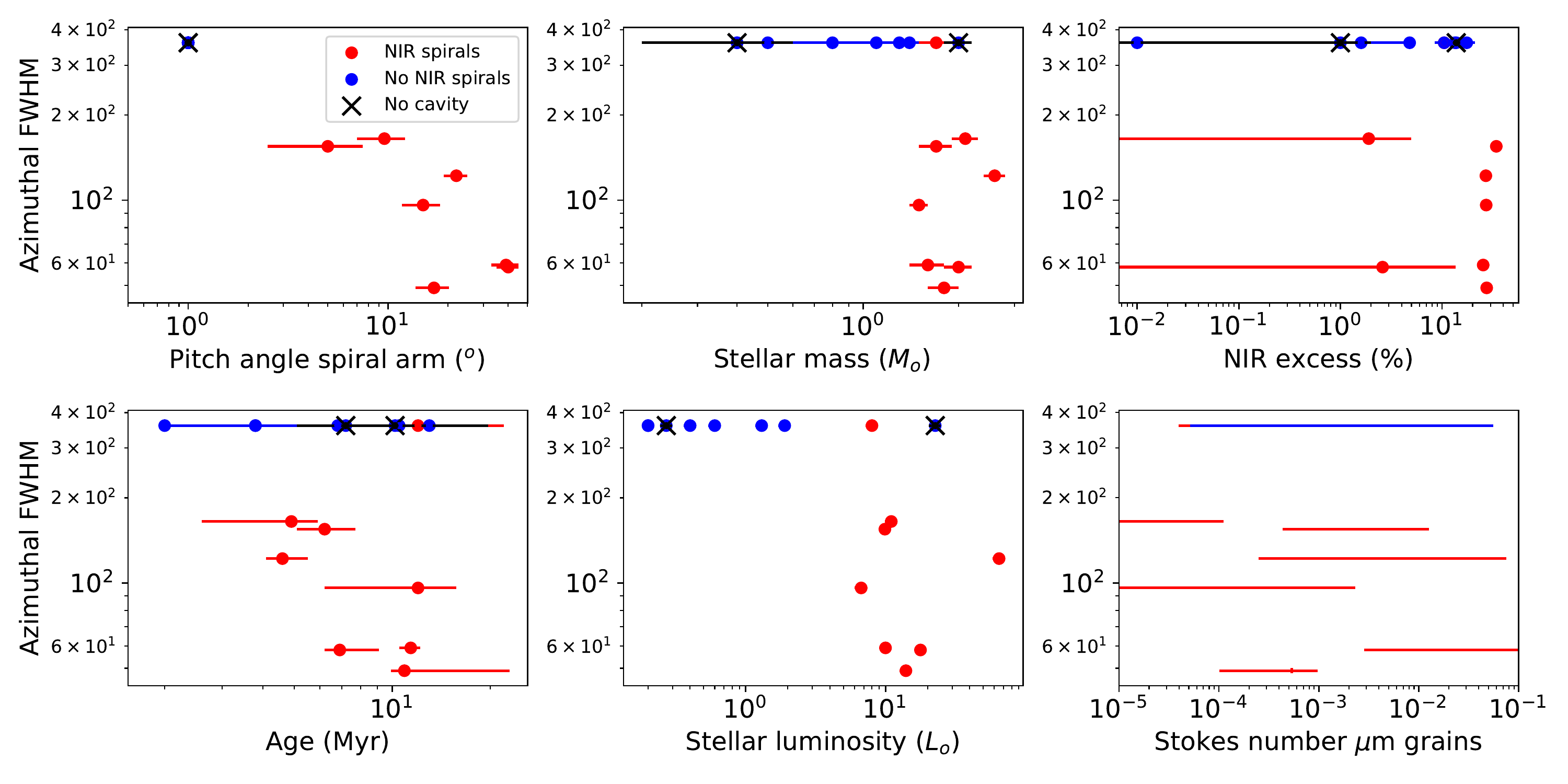}
\caption{Relations between millimeter structure and parameters linked with spiral arms. Red data points are disks with spirals, blue data points are disks without spirals in the NIR.}
\label{fig:stokesspirals}
\end{figure}

Figure \ref{fig:stokesspirals} presents a number of trend plots between the azimuthal extent and parameters that have been linked to spiral arms. The pitch angle was derived from the deprojected scattered light images from the literature (for references see Table \ref{tbl:sample}). When no spiral arms were detected, the pitch angle is set to 1$^{\circ}$. No difference is made between `single' and `double' spirals as the secondary spiral can easily be hidden in part of the disk. Ages, NIR excess and stellar masses are taken from \citet{Garufi2018}. For each plot, we compute the correlation coefficient $r_{\rm corr}$ using the linear regression procedure by \citet{Kelly2007}, resulting in values of $r_{\rm corr}=0.0\pm0.6$ for each plot consistent with a lack of correlation.

Although asymmetries are only seen in disks with spiral arms as already demonstrated by \citet{Garufi2018}, there is no trend between azimuthal extent and the pitch angle. The pitch angle itself depends primarily on the disk temperature (hence aspect ratio) and to a lesser degree on planet mass \citep{Zhu2015,Fung2015}. CQ~Tau's asymmetric nature is debatable but spiral arms have been found. Asymmetries and spiral arms are only seen in stars with stellar masses $>1.5 M_{\odot}$, but not exclusively: HD~163296 is an intermediate mass star without spiral arms and without azimuthal asymmetries. Disks with spirals and asymmetries exist for a range of ages within the sample, although no young disks with spiral arms or asymmetries are known. The sample is intrinsically biased as it only contains disks with ages $\sim$4-10 Myr, which represent a minority disk population that lives longer than the average disk life time of $\sim$3 Myr \citep{Mamajek2009}. The presence of wide gaps and thus dust traps is thought to be the main reason for the longer lifetime, as radial drift is efficiently reduced \citep{Pinilla2020}. NIR excess has been linked before to spiral arms \citep{Garufi2018}, possibly due to changes in the temperature structure in the shadows in the outer disk \citep{Montesinos2016}. Most disks with high NIR excess also show asymmetries, but not for the entire sample.

In the bottom right panel of Figure \ref{fig:stokesspirals} we test if the detection of spiral arms can be linked to the local gas surface density in the same way as the millimeter asymmetries through the Stokes number, following \citet{Veronesi2019}. 
We thus aim to test if the Stokes numbers of $\mu$m-sized dust grains are significantly different in disks with spiral arms and disks with rings in scattered light (see Table \ref{tbl:sample}). In each disk the radial range of the scattered light features (rings and spiral arms) is estimated from the literature, and the Stokes number of a $\mu$m-sized dust grain is computed using the gas surface density profile in that range with equation \ref{eqn:stobs}. The Stokes number of the $\mu$m-sized grains is similar throughout the sample, regardless of contrast or spiral arms. An observed trend would contradict the result of \citet{Veronesi2019} who predicts that rings are only visible at (much) higher Stokes numbers at millimeter wavelengths. The presence of spiral arms might thus be unrelated to the local gas surface density. 

\subsubsection{Link with gaps and stellar properties}

\begin{figure}[!ht]\
\centering
\includegraphics[width=0.5\textwidth]{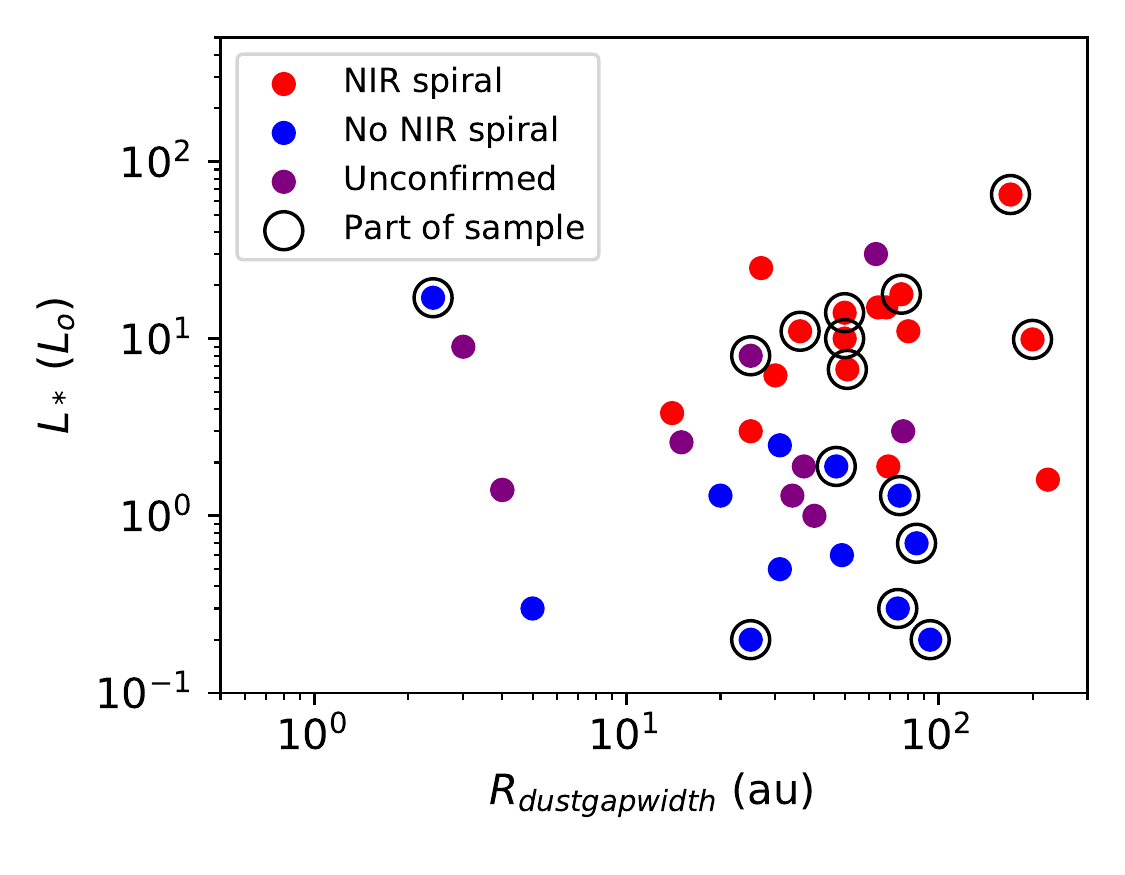}
\caption{Comparison of luminosity, dust gap width and the presence of spirals. The dust gap width represents the inner cavity size for transition disks or the gap width of the most inner gap in ring disks. This plot represents a larger data set including all disks imaged in scattered light from \citet{Garufi2018} for which ALMA data are available in \citet{Francis2020} and \citet{Andrews2018dsharp}. The full table is given in Table \ref{tbl:spiraldata}. The targets from our study are encircled. This plots demonstrates that spirals are only found in disks with high luminosity and a large gap width, as expected from the relation between pitch angle, disk temperature and companion mass.}
\label{fig:lumspirals}
\end{figure}

We explore the link between spirals and stars further in a wider sample comparison in Figure \ref{fig:lumspirals} as more massive stars are generally associated with high luminosities as well. Targets are taken from the scattered light demographics study by \citet{Garufi2018}  for which high resolution ALMA data are available in \citet{Francis2020} and \citet{Andrews2018dsharp} to estimate the dust gap width. $R_{\rm dustgapwidth}$ is defined as $R_{\rm cav}$ for the transition disks \citep{Francis2020} and as the gap width of the inner gap in \citet{Zhang2018dsharp} for the ring disks. The assessment of the presence of spirals is primarily based on the classification by \citet[][Fig. 1]{Garufi2018}, where Spirals and Giants are marked as `spiral' in our sample, Rings and Rims as `no spiral' and Faint, Small or Inclined as `unconfirmed', since the detection of spirals in these disks is hindered by the observational sensitivity, angular resolution, and disk geometry, respectively. Four Giants from \citet{Garufi2018} were  marked by these authors as controversial due to their high inclination, which make the detectability of spiral arms more challenging. These are marked as 'unconfirmed' in our plot. All data are provided in Table \ref{tbl:spiraldata} in the Appendix. 

Figure \ref{fig:lumspirals} shows that all spiral disks lie in the upper right corner of the diagram, with high luminosity and a large gap width, in contrast to non-spirals with either low luminosity or narrow gap width. This difference can be understood as the detectability of spiral arms increases with pitch angle: the pitch angle is correlated with the aspect ratio (disk temperature) and to some extent on companion mass \citep{Dong2015spirals,Fung2015,Zhu2015}. As disk temperature generally scales with stellar luminosity \citep{Dullemond2001} and the gap width roughly with companion mass \citep{Varniere2004}, the pitch angle is thus expected to be larger for more luminous disks with the most massive companions. This link supports a view where the scarcity of spirals around T Tauri stars is due to their low luminosity rather than their young age \citep{Garufi2020}. A larger pitch angle is more easily resolved and thus more likely to be detectable in NIR observations at low inclination. High inclination angles make the detectability of spiral arms more challenging \citep{Dong2016spirals} and indeed, the three purple data points in the upper right corner of Figure \ref{fig:lumspirals} are all disks with inclination $i \sim 40-75^{\circ}$. The two almost face-on disks with tightly wound spirals and a large empty cavity (HD~142527 and GG~Tau) may remain a separate category as these are the only confirmed binaries in this comparison. Another possible connection is that more luminous stars are generally also more massive, and higher mass stars are more likely to have a binary companion \citep{Raghavan2010}, although binary companions have not been detected yet in the majority of these disks (see Section \ref{sct:comps}).

This result demonstrates that spiral arms are possibly present in all disks with gaps, assuming all gaps are opened by planets, but only detected when the planet is sufficiently massive, the star sufficiently luminous and the inclination angle not too high. This explains why spirals are uniquely found in low inclination disks with wide gaps and a high-luminosity star. The thresholds appears to be at $L_*\gtrsim 1.5 L_{\odot}$ and $R_{\rm gapwidth}\gtrsim$15 au, but more data are required to confirm this.

\subsubsection{Link with morphology}
\begin{figure}[!ht]
\includegraphics[width=\textwidth]{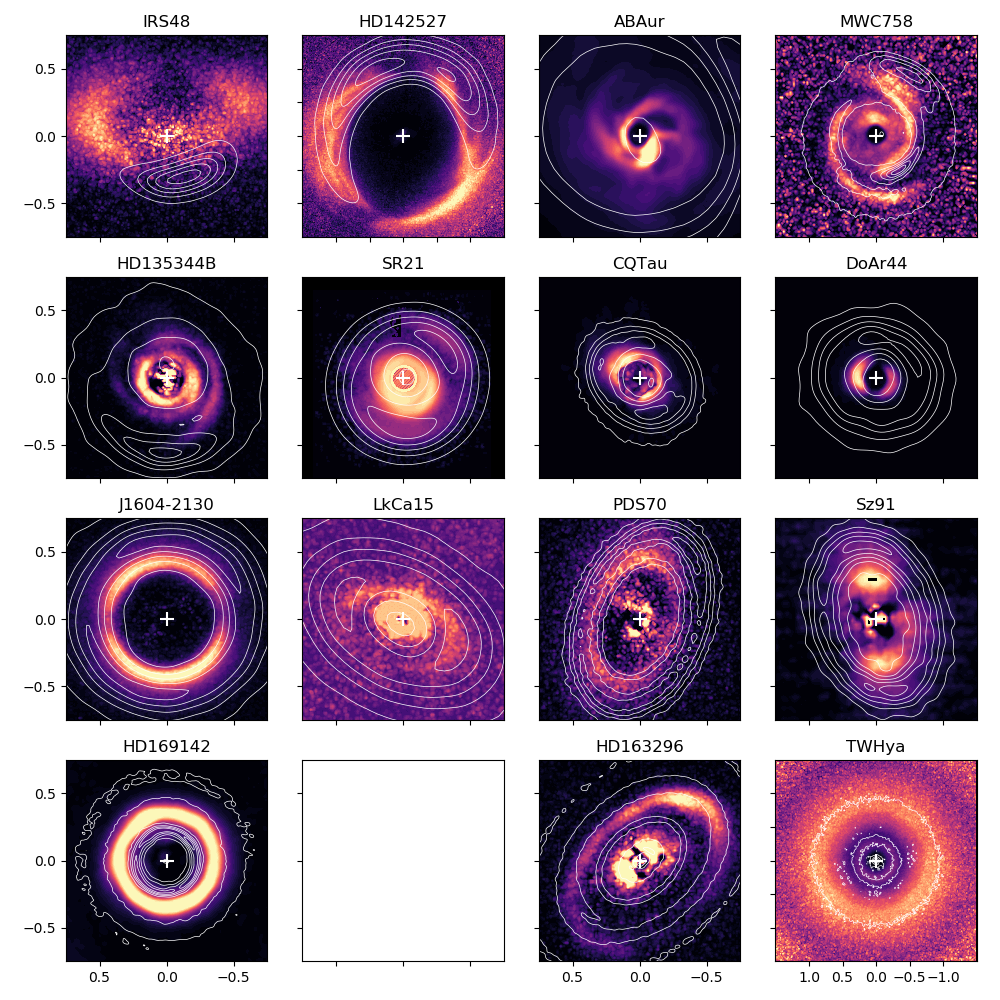}
\caption{Overlay of ALMA millimeter continuum (white contours) onto optical/NIR scattered light images (colors). Most images are from SPHERE/IRDIS, except IRS~48 and CQ~Tau (HiCIAO), HD~142527 and HD~135344B (ZIMPOL) and Sz~91 (NaCo). No NIR image is available for DM~Tau. HD~142527 and TW~Hya are zoomed out compared to the other plots because of their angular size. The bands and references are as follows
IRS~48: H-band \citep{Follette2015}; 
HD~142527: I-band \citep{Avenhaus2017}; 
AB~Aur: H-BAND \citep{Boccaletti2020};
MWC~758: Y-band \citep{Benisty2015};
HD~135344B: I-band \citep{Stolker2016};
SR~21: H-band \citep{Muro-Arena2020};
DoAr~44: H-band \citep{Avenhaus2018};
J1604-2130: J-band \citep{Pinilla2018j1604};
LkCa~15: J-band \citep{Thalmann2016};
PDS~70: J-band \citep{Keppler2018};
CQ~Tau: H-band $r^2$ scaled \citep{Uyama2019};
Sz~91: K-band  \citep{Mauco2019};
HD~169142: J-band \citep{Pohl2017};
HD~163296: H-band \citep{Muro-Arena2018};
TW~Hya: H-band \citep{vanBoekel2016}}
\label{fig:overlays}
\end{figure}

This connection between detectability and luminosity and gap width thus explains the locations of the spiral disks in Figure \ref{fig:stokesspirals} in mass and luminosity, but the link with azimuthal extent remains unclear. Figure \ref{fig:overlays} presents an overlay of scattered light images and ALMA data. The spiral arm(s) and the dust asymmetries appear to be spatially connected, suggesting that the spiral arm and the dust asymmetry may be physically related. Such a comparison has been made before for IRS~48 \citep[][Fig. 12]{Follette2015}, MWC~758 \citep[][Fig. 1c]{Dong2018}, HD~135344B \citep[][Fig. 1b]{Cazzoletti2018}, SR~21 \citep[][Fig. 3d]{Muro-Arena2020} and CQ~Tau \citep[][Fig. 4]{Uyama2019} for the disks in our sample, but also for e.g. V1247 Ori \citep[][Fig. 1b]{Kraus2017} and HD100453 \citep[][Fig. 5]{Rosotti2019}. Proposed scenarios for these connections include e.g. the launching of a spiral by the vortex \citep{vanderMarel2016-spirals} and the detection of part of the spiral in millimeter emission \citep[e.g.][]{Rosotti2019}.

No physical connection is visible in HD~142527 \citep{Avenhaus2017}, AB~Aur \citep{Tang2017}, GG~Tau \citep{Keppler2020} or HD~143006 \citep{Perez2018dsharp}, but this could be related to the limited detectability of the spiral arms in these systems around the radius of the dust asymmetry.

\section{Discussion}
\label{sct:discussion}

\subsection{Diversity asymmetries}
\label{sct:diversity}

In Section \ref{sct:stokes} we compared the azimuthal extents of the dust asymmetries with the local gas surface density through the observational Stokes number, and found a step-like trend which can be matched to a simple dust evolution model with an azimuthal pressure bump in the gas. This result suggests that the diversity of asymmetries and non-asymmetries is not linked to the disk, the companion or a limited lifetime, but to the \emph{local} gas surface density at the location of the pressure bump. This implies that minor azimuthal pressure bumps may be very common in disks, but they are only detected as dust asymmetries when the Stokes number is sufficiently high, i.e. when the local gas surface density is sufficiently low. This also leads to the prediction that dust observations at centimeter wavelengths such as the ngVLA will show a much larger number of asymmetric disks, as a higher observational Stokes number is traced at these wavelengths.

This scenario also explains the existence of dust asymmetries in outer rings, such as seen in e.g. HD135344B (this study), but also in V1247 Ori \citep{Kraus2017} and HD143006 \citep{Andrews2018dsharp} which were not included in this study due to lack of gas analysis. Furthermore, a dust feature identified in high resolution data of TW Hya at 1.3mm at 52 au was interpreted as either a circumplanetary disk or a small azimuthal dust trap \citep{Tsukagoshi2019}. Our data has insufficient sensitivity to reveal this feature. The Stokes number (using our gas surface density profile) is $\sim0.3$ at the location of the feature, which follows the same trend as the other data points in Figure \ref{fig:stokesdata}, and the feature could thus indeed be another azimuthal dust trap. 
As there are no clear correlations between azimuthal extent and typical spiral arm properties (Figure \ref{fig:stokesspirals}), asymmetries may be unrelated to the location or mass of the companion.

If this scenario is correct, asymmetries are not related to the lifetime of a vortex or gas horseshoe. Previous studies have suggested that vortices dissipate on relatively short time scales due to dust feedback \citep{Fu2014,Miranda2017}, although 3D simulations do not reproduce rapid vortex dissipation \citep{Lyra2018}. \citet{Hammer2017} shows that vortices induced by planets may have limited lifetimes when the planet mass growth is not sufficiently fast. Both scenarios have been used to argue that the occurrence rate of asymmetries is caused by the limited lifetime. Our work demonstrates that a time scale may be irrelevant for the occurrence rate. This implies that dissipation of vortices and/or horseshoes could happen on much longer time scales than the lifetime of the disk. 

Figure \ref{fig:stokes} also shows typical values for the gas overdensity consistent with the observations. Several parameters are redundant with each other: a lower viscosity requires a lower gas overdensity and/or gas-to-dust ratio. As vortices are thought to survive only when $\alpha \lesssim 10^{-4}$ \citep{deValborro2007}, this implies that for vortices, the gas-to-dust ratio is likely to be closer to unity, whereas gas horseshoes also survive at high $\alpha$ and may have higher gas-to-dust ratios and/or gas overdensities. Furthermore, the extent dependence becomes shallower for a wider azimuthal width of the gas bump.

Another interesting aspect is the choice of the dust grain size distribution. Both the equilibrium model using a fragmentation velocity and a grain size distribution with a fixed maximum grain size can reproduce the curve. This means that we can neither rule out nor confirm that fragmentation of grains is the limiting effect for dust growth inside the dust trap. Future multi-wavelength data might be able to probe the material properties of these dust particles. In the fragmentation limit the maximum Stokes number should be inversely proportional to $\alpha\cdot T$ \citep{Birnstiel2011}, but the measured asymmetric contrast does not show a clear dependence on the temperature (Figure \ref{fig:stokes}): linear regression analysis results in a $r_{\rm corr}$ coefficient of -0.3 $\pm$0.4.

\subsection{Horseshoe, RWI or spiral?}
\label{sct:origin}

The scenario described in Section \ref{sct:diversity} where dust asymmetries are linked to the local gas surface densities leaves the question of the origin of the gas asymmetry open. Gas horseshoes only appear at the inner edge of a wide, eccentric gap and are unable to trigger secondary asymmetries, so they can be excluded for asymmetries in disks with multiple rings with asymmetries and multiple asymmetries, such as seen in SR~21 and CQ~Tau.. The disks with single dust rings and asymmetric features (HD~142527, IRS~48 and AB~Aur) could still be explained by either gas horseshoe or vortices as the result of Rossby Wave Instability. 

The main distinctions between the gas horseshoe and vortex scenario are the disk viscosity and the mass of the companion: vortices require $\alpha \lesssim 10^{-4}$ to survive for a sufficient amount of time  \citep[e.g.][]{Godon1999,Regaly2012}, whereas the gas horseshoe can exist at higher $\alpha$, and a Rossby Wave Instability occurs at the edge of a planet gap as long as the planet is massive enough to carve a deep gap ($\gtrsim 1 M_{\rm Jup}$) whereas the gas horseshoe requires a mass ratio $q>0.05$, corresponding to $\gtrsim$50-100 $M_{\rm Jup}$ for 1-2 $M_{\odot}$ stellar mass. The Rossby Wave Instability also develops at the edge of an eccentric gap as long as $\alpha \lesssim 10^{-4}$ \citep{Ataiee2013}. The companion in the HD~142527 system has been estimated to be $\sim$150-440 $M_{\rm Jup}$ or $q$=0.07-0.21, consistent with the horseshoe scenario \citep{Price2018}, but in other disks no such companion has been identified.
As the viscosity remains very challenging to constrain observationally, the companion mass provides the best constraint on the origin of the single dust asymmetries. The possible companion masses are discussed in the next Section. 

For the disks with multiple rings and asymmetric features, a vortex remains a likely explanation under the assumption that disks have a viscosity $\alpha \lesssim 10^{-4}$. Hydrodynamic simulations show that the Rossby Wave Instability always develops in these conditions at the edges of gaps and vortices should be very common in disks. Our results demonstrate that the lack of detections of these vortices could simply be due to the local gas surface density.

Another possibility is that the underlying gas asymmetry is in fact part of the spiral density wave, which is supported by the physical connection between spiral arms and mm-dust features discussed in Section \ref{sct:spirals}. The reason that only a small part of the spiral arm is visible in the millimeter, is that millimeter grains are only present at the edge of the planet gap, where the spiral density wave can lead to a further concentration of the dust azimuthally, but again, only when the local radial gas surface density is low enough. Small mismatches between the curve of the mm-dust and scattered light features such as seen in HD~135344B \citep{vanderMarel2016-spirals} can be understood as the emission originates from different heights in the disk \citep{Rosotti2019}. 

Spiral waves launched by a planet are rotating with respect to the background gas flow: they run over the dust particles with little time for the particles to react. Dust particles thus cannot get trapped and get carried along in spiral density waves: the time scales for dust accumulation in dust traps are at least 100 times longer than the local orbital period \citep{Birnstiel2013}. However, spiral waves still lead to changes in the pressure scale height and vertical flows, which may also lead to a different spatial distributions of different particle sizes that reproduces the observed morphologies. Whether asymmetric features in the millimeter continuum really can be part of a spiral density wave  remains a question.

Furthermore, it is unclear whether the disks without clear physical connection between spiral arm and mm-dust feature listed above are intrinsically different (e.g. because of a substellar rather than a planetary companion) or whether these large scale asymmetries are perhaps connected to the spiral arm after all. Spiral arms traced in high resolution $^{12}$CO observations \citep[such as seen on larger scale in e.g. HD142527][]{Christiaens2014} might help to reveal this connection as $^{12}$CO remains visible at lower surface densities than small dust grains. 

Multi-epoch observations of asymmetric mm-dust features in disks are required to measure their rotational speed, in order to see if they move along with the spiral (with the orbital speed of the companion) or on their own Keplerian orbit. The latter would directly rule out trapping in a spiral density wave.  

\subsection{Implications for companions}
\label{sct:discussioncomps}

Figure \ref{fig:companions} shows the known limits for companions in each of the disks. The 3 confirmed companions (HD~142527B, PDS70b and PDS70c) and 3 companion candidates (MWC758b, LkCa15b and HD169142b) are located at or around the gas gap radii well inside the dust ring radii. The HD~142527B companion was detected at a small separation of 12 au, but orbital fitting indicates the orbit is highly eccentric, and the companion may reach a separation of at least 57 au at apoapsis \citep{Claudi2019}, very close to the derived gas gap radius. For the 10 systems where limits are derived through a contrast curve, limits are known at the gas gap location for all except 1 system (CQ~Tau), and for SR~21 the limits are very marginal at the gap location.  Only for AB~Aur no limits on companions exist. If the gap radii are overestimated and the gap radius is even closer in to the star, the contrast curves only provide limits for about half of the sample.

\begin{table}[!ht]
\centering
\caption{Possible companion mass at gap radius}
\label{tbl:possiblecompanions}
\begin{tabular}{llllllll}
\hline
Target&$R_{\rm gap}$&$M_p$&$q$&$R_{\rm p,c}$&$M_{p,c}$&$q_{p,c}$&Gap edge\\
&(au)&($M_{\rm Jup}$)&&(au)&($M_{\rm Jup}$)&&\\
&(1)&(2)&(3)&(4)&(5)&(6)&\\
\hline
IRS48&22&$\lesssim$100&$\lesssim$0.05&-&-&-&RWI/horseshoe\\
HD142527&65 &$\lesssim$9.5&$\lesssim$0.005&18-57&270$\pm$157&0.15$\pm$0.08&horseshoe\\
ABAur&64&-&-&\footnote{There is an estimate of a 4-13 $M_{\rm Jup}$ at 30 au from the spiral twist in \citet{Boccaletti2020}, but we leave this out due to its uncertain nature.}&-&-&RWI/horseshoe\\
MWC758&22&$<$38&$<$0.02&20&52$\pm$10&0.03$\pm$0.003&RWI/spiral+RWI/spiral\\
HD135344B&15&$<$60&$<$0.04&-&-&-&ring+RWI/spiral\\
SR21&7&$<$352&$<$0.16&-&-&-&RWI/spiral\\
CQTau&13&-&-&-&-&-&ring/RWI/spiral\\
DoAr44&14&$<$44&$<$0.03&-&-&-&ring\\
J1604-2130&37&$<$2.4&$<$0.002&-&-&-&ring\\
LkCa15&25&$<$14&$<$0.01&15&1-15&0.001-0.01&ring\\
PDS70&23&-&-&22, 35&5-9, 3.3-5.5&0.004-0.01&ring\\
Sz91&29&$<$14&$<$0.03&-&-&-&ring\\
HD169142&12&-&-&11&30$\pm$2&0.017$\pm$0.001&ring\\
&42&$<$14&$<$0.008&38&3&0.002&ring\\
DMTau&12$<$7&$<$0.02&-&-&-&-&ring\\
HD163296&-&-&-&-&-&-&ring\\
TWHya&-&-&-&-&-&-&ring\\
\hline
\end{tabular}\\
Explanation columns: (1) Estimated gap radius from the $^{13}$CO profile; (2) Maximum companion mass at $R_{\rm gap}$ according to the contrast curve; (3) Maximum mass ratio at $R_{\rm gap}$ according to the contrast curve; (4) Radius of detected companion candidate(s) from direct imaging; (5) Companion candidate mass estimate from direct imaging; (6) Companion candidate mass ratio estimate from direct imaging.
\end{table}

The contrast curves, which have been derived using hot-start models, rule out mass ratios $q>$0.05 ($M_p>50 M_{\rm Jup}$, the minimum threshold for the formation of gas horseshoes) in the targeted regimes. Table \ref{tbl:possiblecompanions} provides the limits for the possible companions and expected structure at the gap edge. However, high mass ratios are still possible in the inner parts of the disk where no contrast could be measured, which is particularly relevant for the disks where the derived gas gap radius is not covered by the contrast curve, such as CQ~Tau, SR~21 and AB~Aur, and perhaps IRS~48. Such a high mass ratio was recently suggested for IRS~48 and AB~Aur for reproducing the CO kinematics and dust contrast through a circumbinary simulation \citep{Calcino2019,Poblete2020}. 

The large derived ratios between the gas gap radius and dust ring radius from Figure \ref{fig:gapratio} imply minimum companion masses $>15 M_{\rm Jup}$ which are in the brown dwarf and stellar regime, for the assumption that the gaps are cleared by a single companion on a circular orbit. On the other hand, the contrast curves generally rule out companion masses $>50 M_{\rm Jup}$ at the gap location. This limits the companion masses (at $R_{\rm gap}$) at the gap location to the brown dwarf regime. This would be consistent with the derived companion candidate masses in MWC~758 and HD~169142, although these masses remain highly uncertain due to the lack of available data for analysis of the contributions by a circumplanetary disk, if present. 

On the other hand, it is very likely that simple planet-disk interaction models with a fixed orbit such as those used by \citet{Facchini2017gaps} are insufficient to derive planet masses from CO vs dust images. A disk gap is thought to become eccentric when the mass ratio $q\gtrsim0.003$ due to eccentric Lindblad resonances \citep{KleyDirksen2006}; the back reaction the disc exerts on the companion has been shown to grow its orbital eccentricity \citep[e.g.][]{Papaloizou2001,Ragusa2018}, which is efficient down to Super-Jovian planet masses when accretion onto the planet is included \citep{dAngelo2006}. \citet{Muley2019} demonstrated that a proper planet-disk interaction model with a single planet, including accretion onto the planet and migration, is able to reproduce the observed gas gap and dust ring in PDS70 with a planet with a mass of $\sim4 M_{\rm Jup}$ after 4 Myr, with a natural eccentricity growth up to $e\sim0.3$. This simulation was run to explain PDS70 with only PDS70b, as PDS70c was still unknown at the time. This scenario has recently been proposed to explain MWC~758 as well \citep{Calcino2020}.

We hypothesize that the wide transition disk cavities in our sample are also caused by eccentric, Super-Jovian planets: these planets are in the 3-15 $M_{\rm Jup}$ regime and their eccentric orbits have developed naturally as shown in \citet{Muley2019}. Due to the eccentric planet orbit, the disk may no longer appear eccentric. Eccentric disk cavities have been observed in MWC~758 \citep{Dong2018} and AB~Aur (this work), but are generally hard to determine observationally. The main motivation for this scenario is thus the large separation between the dust cavity radius and the deduced gas gap, which is thought to be representative of the companion orbit. Note that also brown dwarfs are expected to carve eccentric gaps considering their high mass ratios. It is also possible that multiple companions (such as seen in PDS70) are responsible for the wide gaps. The (sub)stellar companions required for gas horseshoes are ruled out in the majority of our disks, with the exception of IRS~48, HD~142527, AB~Aur and CQ~Tau.

The occurrence rate of massive companions from direct imaging surveys in older systems argues against brown dwarf companions as a common explanation for transition disks. Super-Jovians (5-13 $M_{\rm Jup}$) have an occurrence rate of 8.9\% at 10-100 au for intermediate mass (1.5-5 $M_{\odot}$) stars \citep{Nielsen2019}, which is the stellar mass range of the majority of our sample. Above that mass treshold, the occurrence rate of brown dwarfs (13-80 $M_{\rm Jup}$) at wide orbits is much lower \citep[$\sim$1\%][also known as the 'brown dwarf desert']{Nielsen2019}, but the occurrence of stellar companions ($>80 M_{\rm Jup}$) or binarity rate at 10-100 au is again increased, with a fraction of $\sim$15\% in the 1-2 $M_{\odot}$ stellar mass range \citep{Moe2019}. This suggests that transition disks are more likely caused by Super-Jovians or stellar companions than brown dwarfs.

The narrow gaps in HD~163296 and TW~Hya are consistent with lower mass planets $<5 M_{\rm Jup}$ for which eccentricity is unlikely to develop. This is possibly a distinction between so-called transition disks and ring disks  \citep{vanderMarel2019}: wide gaps only develop when the planet is sufficiently massive to develop an eccentricity, which requires $q>$0.003. A ring disk may host multiple, lower-mass companions.

\section{Conclusions}
\label{sct:conclusions}

To summarize our study, we conclude that:
\begin{itemize}
\item Asymmetries in the dust appear to be linked to a low local gas surface density through the observational Stokes number.
\item Current dust and gas observations cannot distinguish between vortices (caused by planetary companions) and gas horseshoes (caused by (sub)stellar companions with a mass ratio $q>$0.05) in single-ring systems.
\item The link between the presence of asymmetries and local gas surface density can explain why asymmetric features in multi-ring systems are always seen in the outer ring. 
\item The underlying gas asymmetries in multi ring systems could be linked to either vortices (if $\alpha \lesssim 10^{-4}$ in disks) or spiral arms. The latter could explain the observed correlation between the presence of spiral arms and asymmetries in disks \citep{Garufi2018}. 
\item The diversity in asymmetries does not require dissipation of vortices or gas horseshoes and their lifetimes may be much longer than previously thought.
\item Current direct imaging results are consistent with Super-Jovian and substellar companions at orbits well inside the dust rings as the cause of large gaps in transition disks.
\item The ratios between the dust ring radii and gas gap radii suggest that either Super-Jovian (3-15 $M_{\rm Jup}$) companions on naturally occurring eccentric orbits or (sub)stellar ($>$15 $M_{\rm Jup}$) on circular orbits would be responsible for the wide gaps. (Sub)stellar companions ($q>$0.05 or $>$50 $M_{\rm Jup}$) are ruled out by contrast curves for the majority of the sample at the gap location, but remain possible for some disks at even smaller radii. 
\item The detection of spiral arms in scattered light images is linked to high luminosity stars with wide gaps, which can be understood in terms of the pitch angle which depends on disk temperature and companion mass. This can also explain the scarcity of detected spiral arms around T Tauri stars. 
\end{itemize}
Our results predict that dust observations at centimeter wavelengths such as the ngVLA will show a much larger number of asymmetric features. Further studies of asymmetric and spiral features due to companions for a large grid of models to obtain predictions and observables are required to fully disentangle the origin and diversity of these feature in observational data.

  \begin{acknowledgements}
  \emph{Acknowledgements.} The authors would like to thank the referee for their constructive report, which has greatly improved the clarity of the manuscript. They would also like to thank Wlad Lyra and Thayne Currie for useful discussions, and Yann Boehler, Satoshi Mayama, Davide Fedele, Miriam Keppler, Maria Ubeira Gabellini, Paola Pinilla, Anthony Boccaletti, Taichi Uyama, Tomoyuki Kudo, Tomas Stolker, Myriam Benisty, Roy van Boekel, Gabriela Muro-Arena and the SPHERE consortium for providing their data fits files. This study was initiated by discussions at the Great Barriers conference in July 2019 in Queensland, Australia. N.M. acknowledges support from the Banting Postdoctoral Fellowships program, administered by the Government of Canada. T.B. acknowledges funding from the European Research Council (ERC) under the European Union’s Horizon 2020 research and innovation programme under grant agreement No 714769 and funding from the Deutsche Forschungsgemeinschaft under Ref. no. FOR 2634/1 and under Germanys Excellence Strategy (EXC-2094–390783311). E.R. acknowledges financial support from the European Research Council (ERC) under the European Union's Horizon 2020 research and innovation programme (grant agreement No 681601). D.P. and V.C. acknowledge funding from the Australian Reasearch Council via DP180104235.

This paper makes use of the
  following ALMA data:  
2011.0.00318.S,
2011.0.00465.S,
2012.1.00158.S,
2012.1.00303.S,
2012.1.00422.S,
2012.1.00631.S,
2012.1.00725.S,
2012.1.00761.S,
2012.1.00870.S,
2013.1.00100.S,
2013.1.00498.S,
2013.1.00592.S,
2013.1.00601.S,
2013.1.01020.S,
2015.1.00678.S,
2015.1.00686.S,
2015.1.00888.S,
2016.1.00340.S,
2016.1.00344.S,
2016.1.00484.L,
2016.1.00629.S,
2017.1.00492.S,
2017.1.01404.S,
2017.1.01460.S,
2017.A.00006.S.
 ALMA is a partnership of ESO (representing its member states), NSF (USA) and
  NINS (Japan), together with NRC (Canada) and NSC and ASIAA (Taiwan),
  in cooperation with the Republic of Chile. The Joint ALMA
  Observatory is operated by ESO, AUI/NRAO and NAOJ.  
  \end{acknowledgements}

\bibliographystyle{aasjournal}

\appendix

\section{Visibility analysis}
\label{sct:viscurves}
In this section, we present the visibility analysis of the asymmetric disks in order to compute the FWHM of each asymmetric feature. The asymmetric disks are fit to a 2D profile $I(r,\phi)$, describing the radial and azimuthal features as Gaussians. The fitting is performed in the visibility plane comparing both the Real and Imaginary components, using the \texttt{galario} tool set to Fourier transform and sample the model \citep{Tazzari2018}.

The disks can be described by a combination of one or two rings and asymmetries. HD~142527 and AB~Aur is best fit with a combination of a ring and an asymmetry. This model is parametrized as follows:

\begin{equation}
I(r,\phi) = I_1 e^{(\frac{-(r-r_{c1})^2}{2r_{w1}^2})}e^{(\frac{-(\phi-\phi_{c1})^2}{2\phi_{w1}^2})} + I_2 e^{(\frac{-(r-r_{c2})^2}{2r_{w2}^2})}e^{(\frac{-(\phi-\phi_c)^2}{2\phi_w^2})} \\ + I_3 e^{(\frac{-(r-r_c3)^2}{2r_w3^2})} + I_4 e^{(\frac{-(r-r_c4)^2}{2r_w4^2})}
\end{equation}

IRS~48 can be described by a single 2D Gaussian, where the radial profile is found to be best fit with a 4th power rather than a second power:
\begin{equation}
I(r,\phi) = I_1 e^{(\frac{-(r-r_c)^4}{2r_w^4})}e^{(\frac{-(\phi-\phi_c)^2}{2\phi_w^2})}
\end{equation}

\begin{figure}[!ht]
\centering
\includegraphics[width=\textwidth]{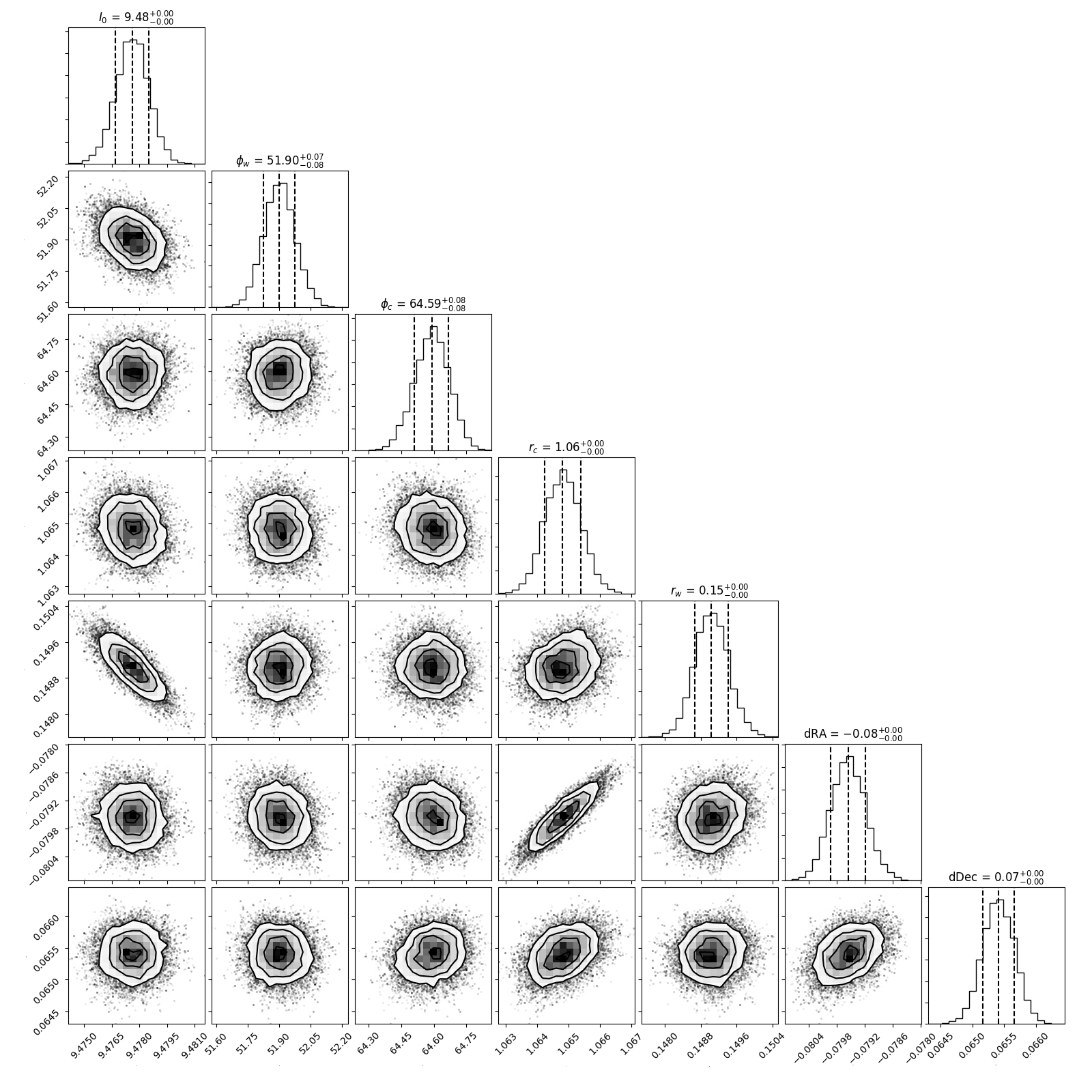}
\caption{Example of the MCMC results for the asymmetry in AB~Aur, showing the two-dimensional posterior distributions for the MCMC fit. The median values and 1$\sigma$ standard deviation of the best-fitting parameters are indicated by dashed lines.}
\label{fig:mcmc}
\end{figure}

The best-fit parameters are found by careful exploration of the parameter values for initial estimates, followed by a fit to the visibilities with the Markov-chain Monte Carlo (MCMC) code \texttt{emcee} \citep{Foreman2013} for constraining each asymmetry. In the MCMC, 70 walkers were used in combination with 2000 steps. The position angle and inclination are fixed, taken from Table \ref{tbl:sample}. The phase center is fit as well and listed in the table. The MCMC runs converge to gaussians distributions with small statistical errors (see example in Figure \ref{fig:mcmc}). In particular, the statistical errors on $\phi_w$ (the parameter of interest) are less than 1$^{\circ}$. The offsets at long baselines in the imaginary curve of SR~21 have low weights and don't contribute much to the fit. For HD~142527 it was not possible to find convergence, likely due to the complexity of the shadow around the peak due to the misaligned inner disk \citep{Casassus2018}. The best-fit parameters are thus not as well constrained as the other disks, but sufficient for our purposes. 

\begin{figure}[!ht]
\includegraphics[width=0.3\textwidth]{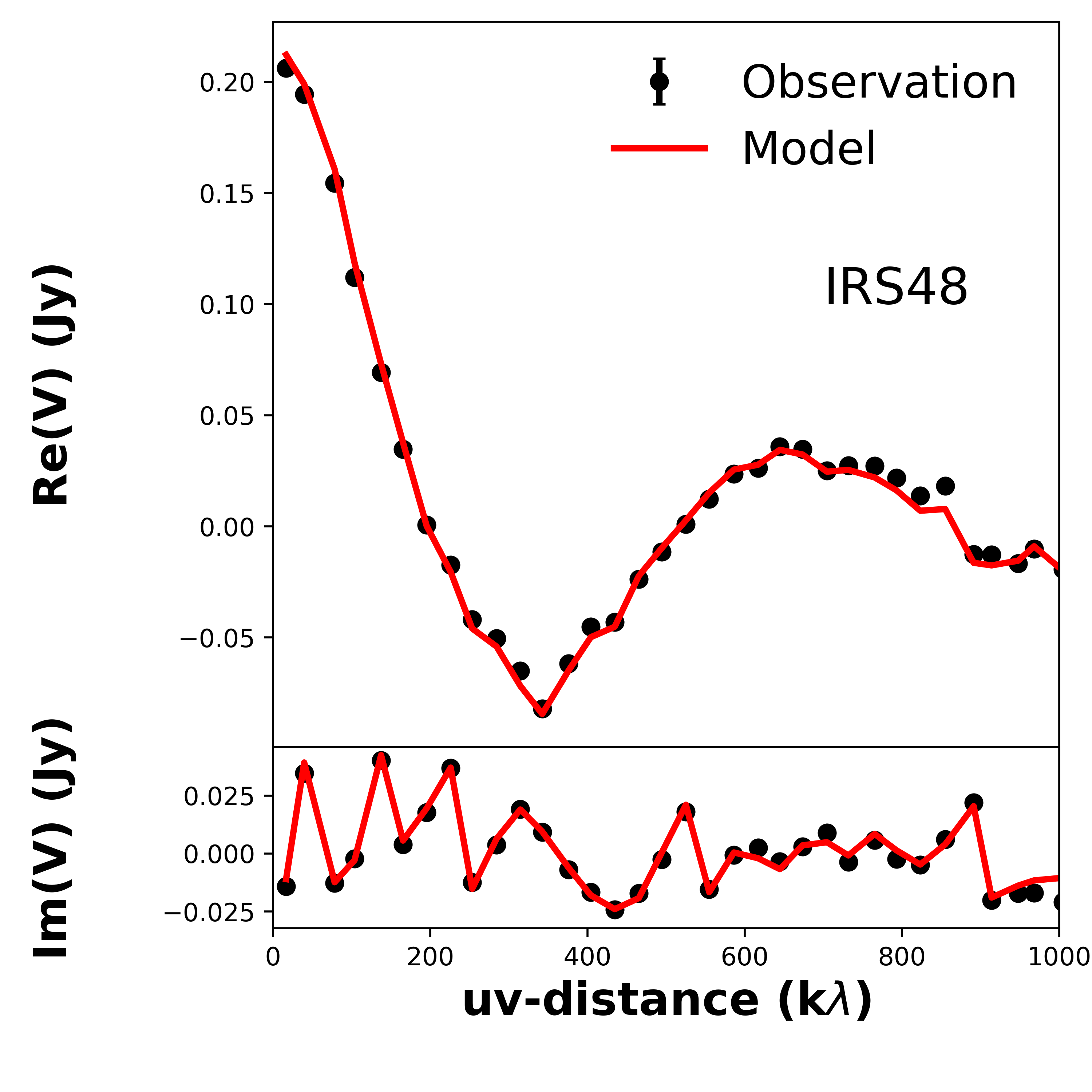}
\includegraphics[width=0.3\textwidth]{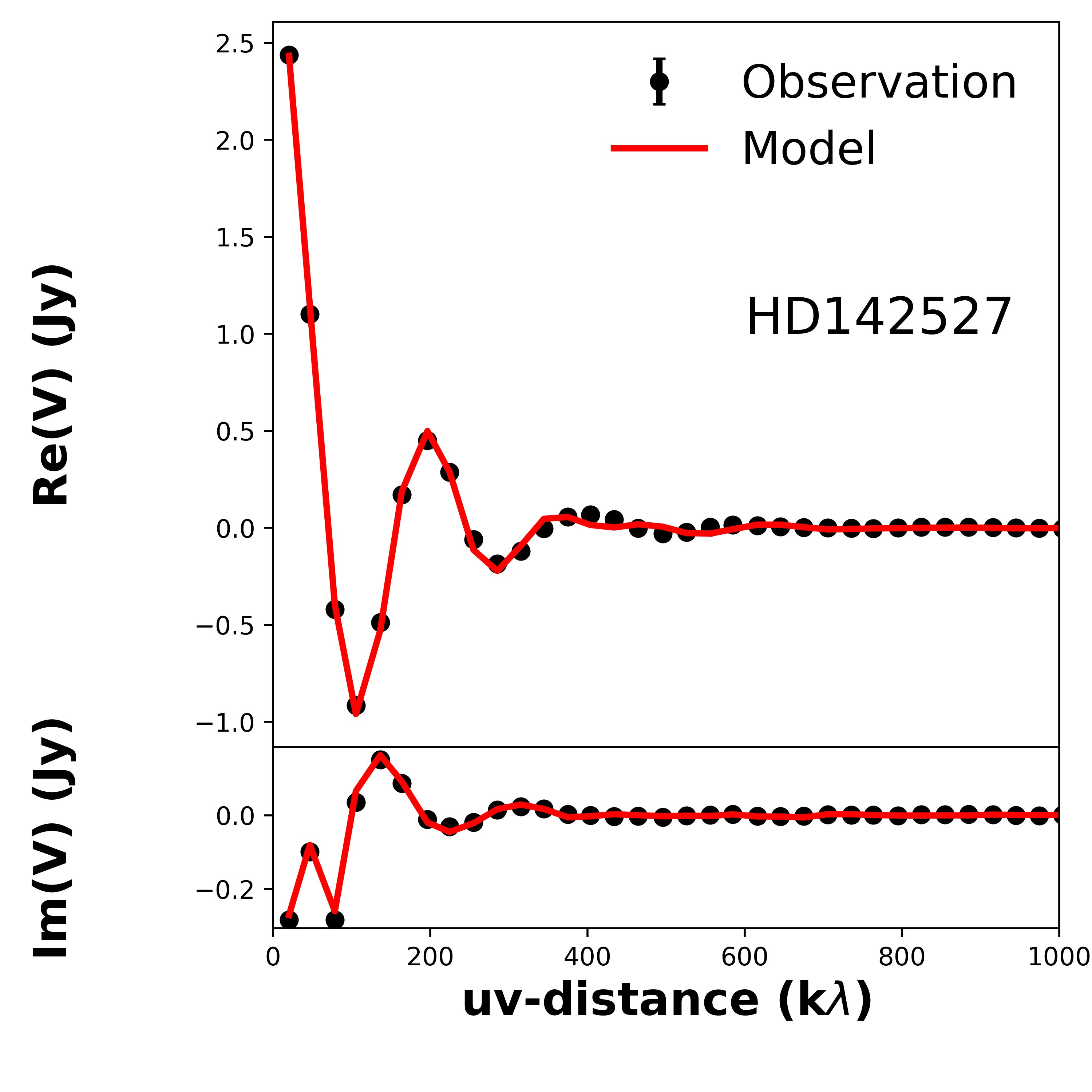}
\includegraphics[width=0.3\textwidth]{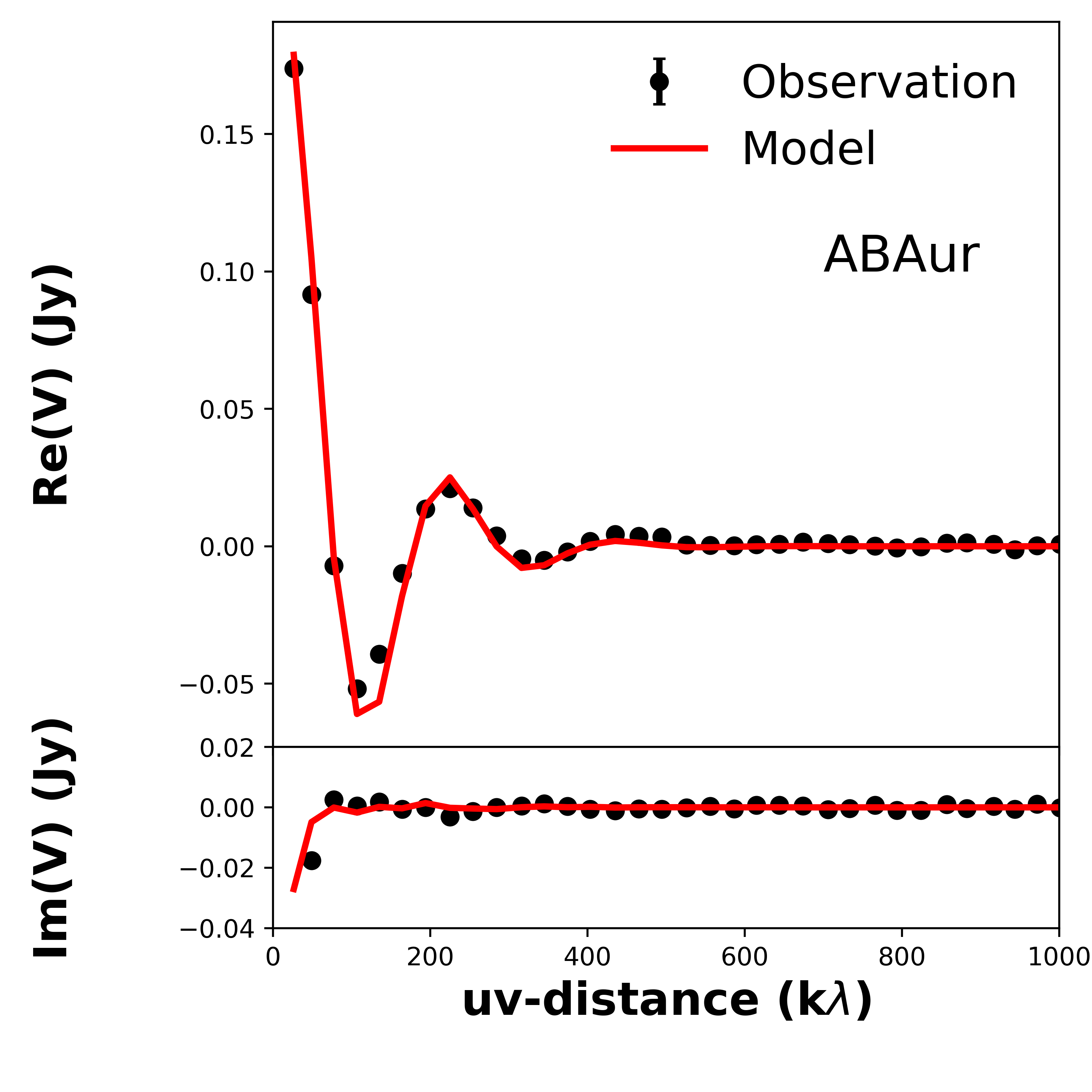} \\
\includegraphics[width=0.3\textwidth]{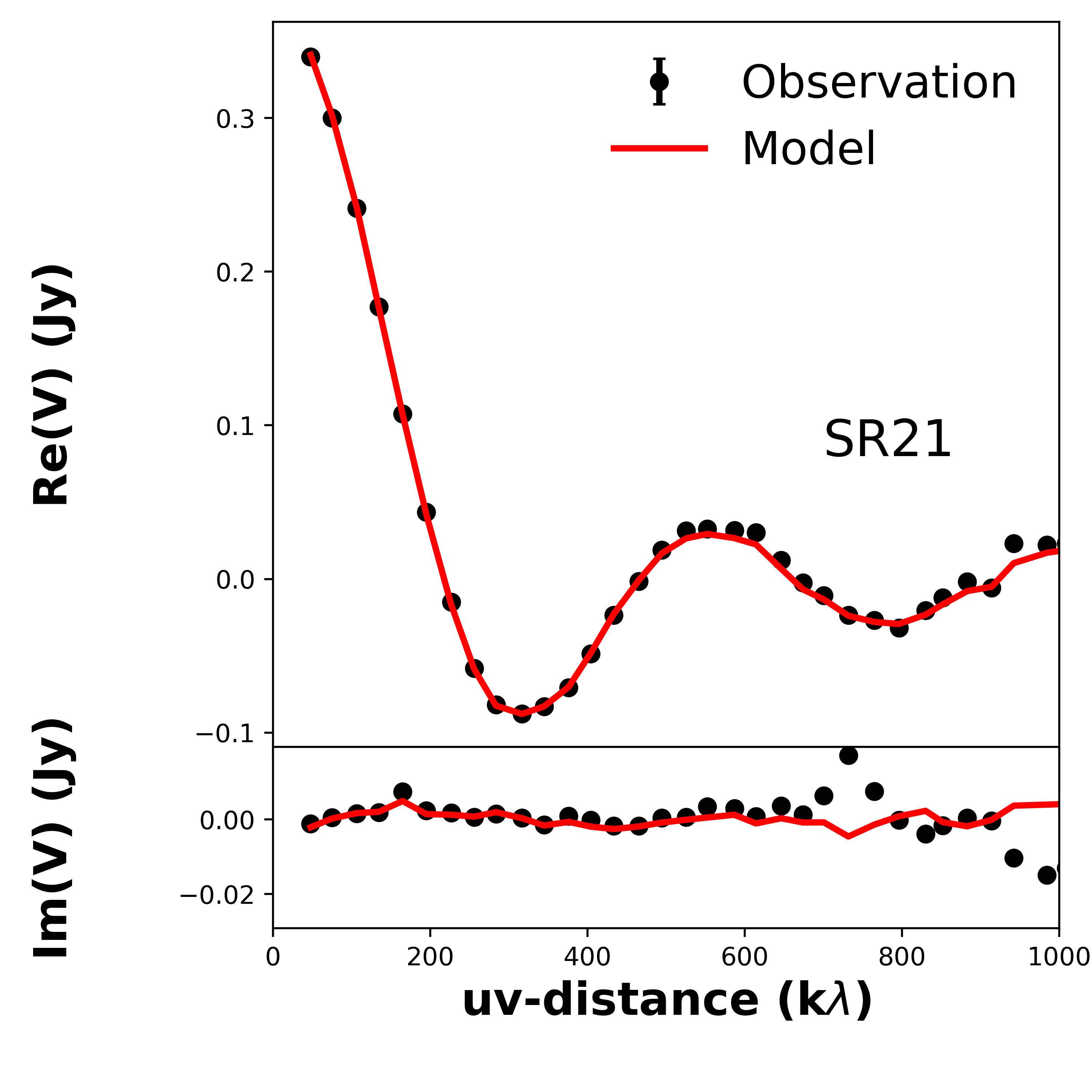}
\includegraphics[width=0.3\textwidth]{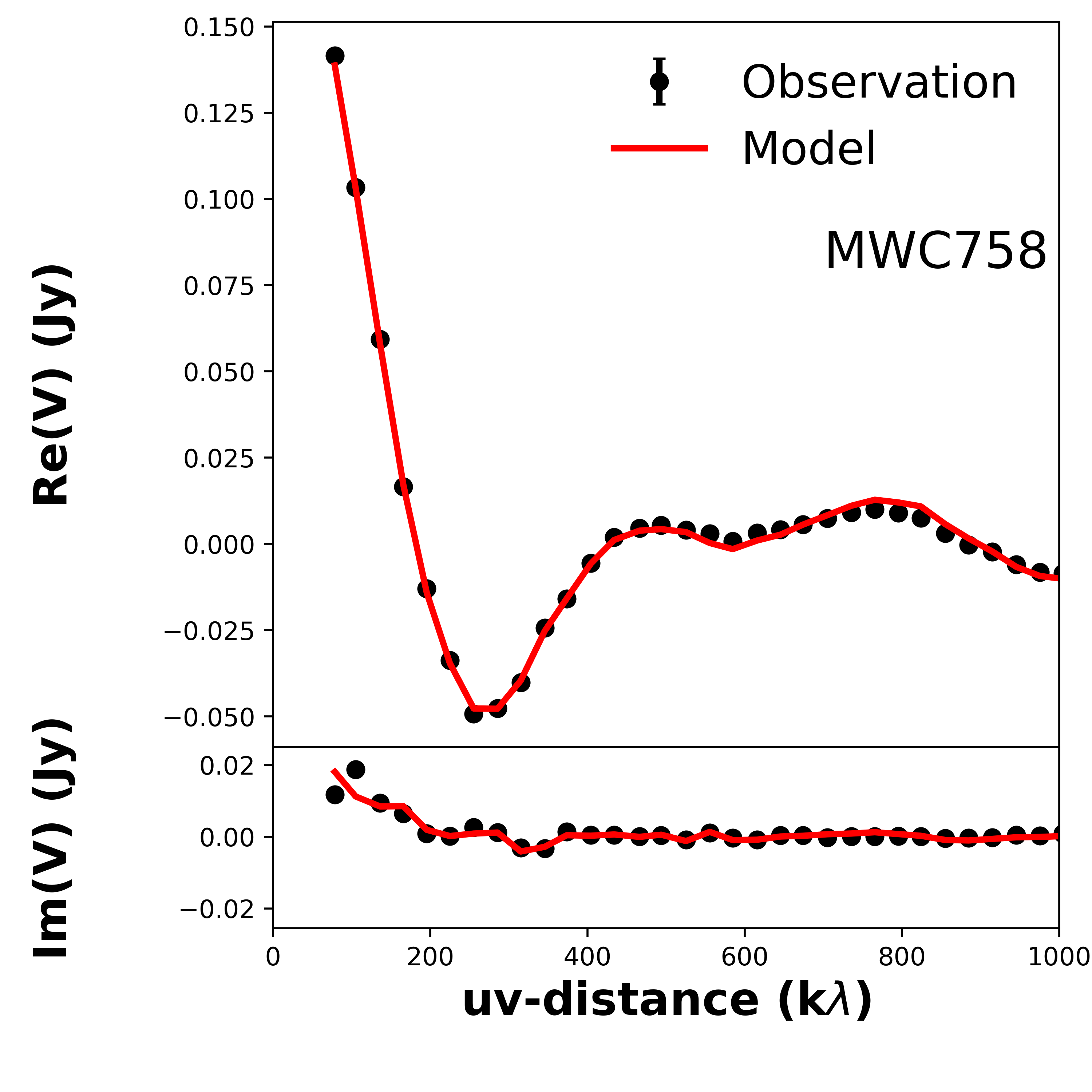}
\includegraphics[width=0.3\textwidth]{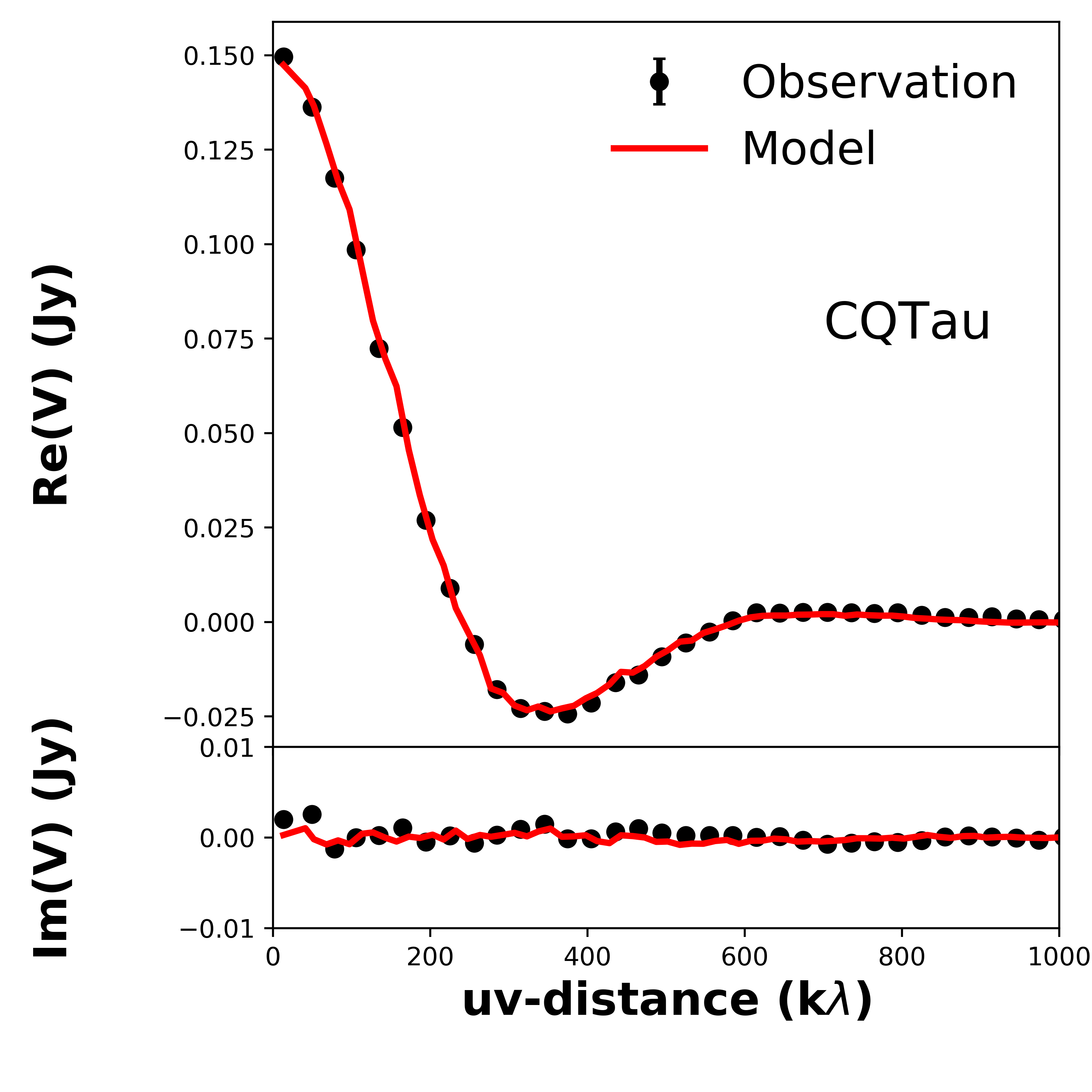}
\caption{Visibility curves and best-fit models (in red) of each of our asymmetric disks.}
\label{fig:fitvisibilities}
\end{figure}

\begin{figure}[!ht]
\includegraphics[width=0.5\textwidth]{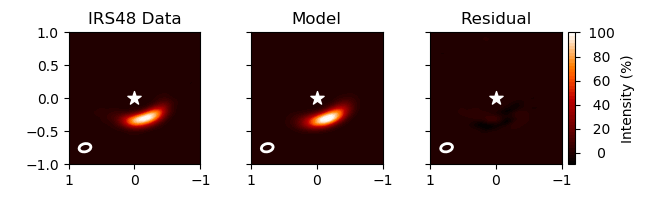}
\includegraphics[width=0.5\textwidth]{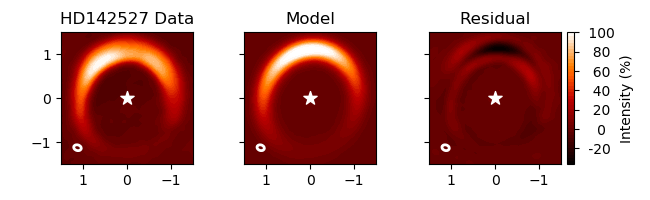}\\
\includegraphics[width=0.5\textwidth]{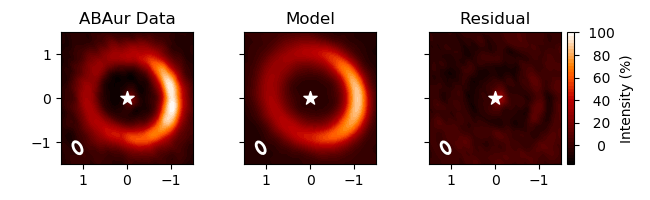} 
\includegraphics[width=0.5\textwidth]{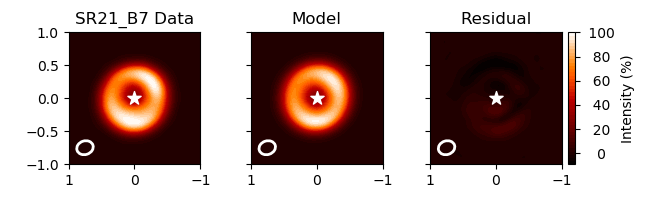}\\
\includegraphics[width=0.5\textwidth]{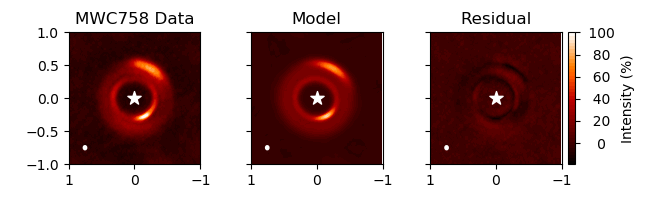}
\includegraphics[width=0.5\textwidth]{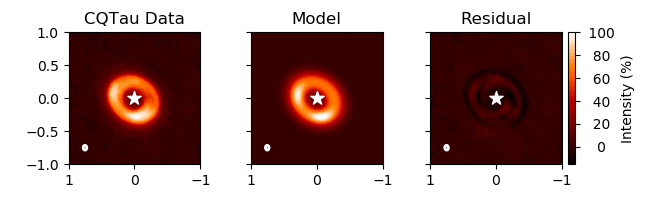}
\caption{Best fit models from the visibility analysis, normalized to the peak of the original image.}
\label{fig:fitimages}
\end{figure}

The best-fit models are shown in the visibility curves in Figure \ref{fig:fitvisibilities}. The best-fit models are mapped onto the observed visibilities, imaged and subtracted to image the residuals. This comparison is presented in Figure \ref{fig:fitimages}. The residuals contain typically 12-22\% of the peak image, similar to the best fits of \citet{Cazzoletti2018} for HD135344B. These residuals are due to the structures that cannot be well represented by a simple double Gaussian as used to model the asymmetry. The best-fit parameter values are listed in Table \ref{tbl:fitparameters}.

\begin{table}[!ht]
\centering
\caption{Best fit parameters of asymmetric models}
\label{tbl:fitparameters}
\begin{tabular}{l|cccccc}
\hline
Parameter&IRS~48&HD~142527&AB~Aur&SR~21&MWC~758&CQ~Tau\\
\hline
$\log I_1$ (Jy/sr)	&11.09	&10.92	&9.48	&10.35	&10.60	&9.75\\
$r_{c1}$ (")	&0.49	&1.15	&1.06	&0.40	&0.31	&0.30\\
$r_{w1}$ (")	&0.10	&0.22	&0.25	&0.06	&0.02	&0.05\\
$\phi_{c1}$ ($^{\circ}$)	&165	&240	&65	&340	&123	&245\\
$\phi_{w1}$ ($^{\circ}$)	&25	&66	&52	&35	&21	&24\\
$\log I_2$ (Jy/sr)	&-	&-	&-	&10.35	&10.55	&9.90\\
$r_{c2}$ (")	&-	&-	&-	&0.42	&0.57	&0.31\\
$r_{w2}$ (")	&-	&-	&-	&0.06	&0.04	&0.05\\
$\phi_{c2}$ ($^{\circ}$)	&-	&-	&-	&130	&360	&132\\
$\phi_{w2}$ ($^{\circ}$)	&-	&-	&-	&70	&20	&25\\
$\log I_3$ (Jy/sr)	&-	&-	&9.25	&10.30	&10.00&10.10\\
$r_{c3}$ (")	&-	&-	&1.00	&0.39	&0.32	&0.30\\
$r_{w3}$ (")	&-	&-	&0.25	&0.06	&0.03	&0.12\\
$\log I_4$ (Jy/sr)	&-	&-	&-	&10.45	&9.85	&-\\
$r_{c4}$ (")	&-	&-	&-	&0.21	&0.45	&-\\
$r_{w4}$ (")	&-	&-	&-	&0.03	&0.10	&-\\

\hline
RA&16:27:37.182&15:56:41.872&04:55:45.863&16:27:10.27&05:30:27.537&05:35:58.471 \\
Dec&-24:30:35.38&-42:19:23.655&+30:33:03.985&-24:19:13.068&+25:19:56.583&+24:44:53.614\\
\hline
\end{tabular}
\end{table}

The $\phi_w$ value provides the estimate for the FWHM of each asymmetry by multiplication with 2.36. 


\clearpage
\newpage

\section{$^{13}$CO maps}
In this section we present the integrated intensity of the $^{13}$CO data of each of our targets. 

\begin{figure}[!ht]
\includegraphics[width=\textwidth]{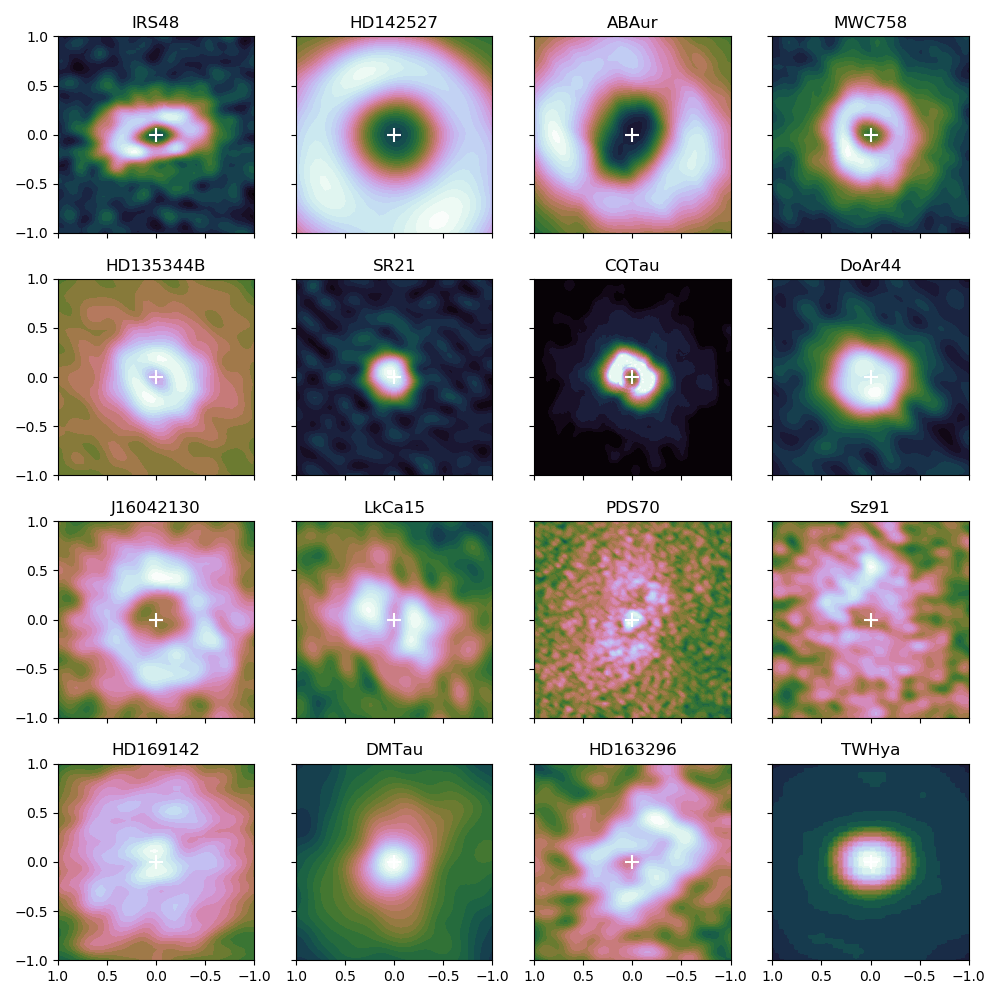}
\caption{Integrated intensity maps of $^{13}$CO of each target (see also Figure \ref{fig:COgaps}).}
\label{fig:COimages}
\end{figure}

\section{Kinematics}
\label{sct:kinematics}
\begin{figure}[!ht]
\includegraphics[width=\textwidth]{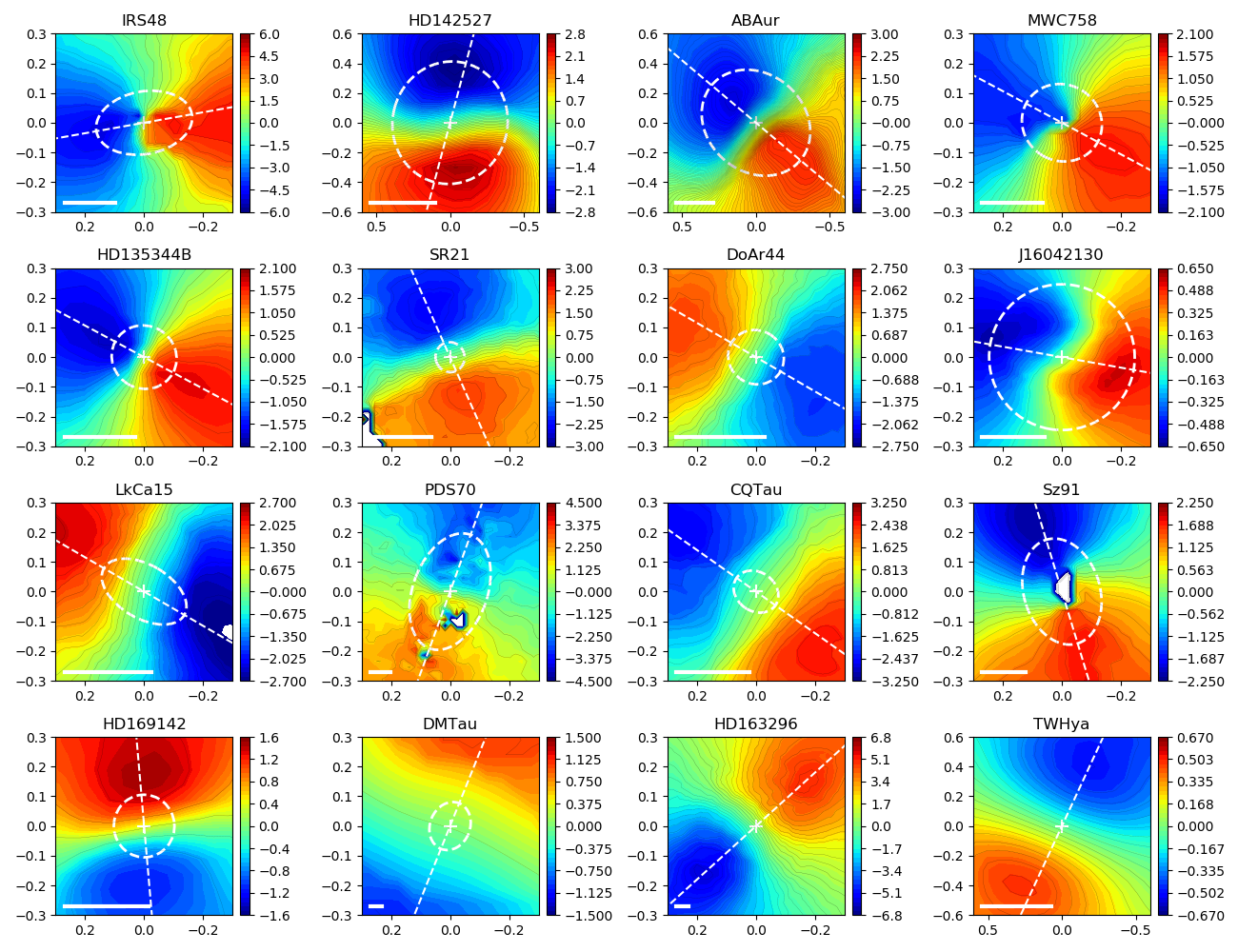}
\caption{First moment map of each target. The colors show the gradients of the velocity, and the grey contours indicate the velocity resolution of the observations (see Table \ref{tbl:kinematics}). The white dashed ellipses mark the gas gap radii and the diagonal white dashed line indicates the position angle of the outer disk to guide the eye. The images are zoomed into the central part of each disk. The horizontal bar at the bottom shows the mean diameter of the beam.}
\label{fig:kinematics}
\end{figure}

We compare the kinematics of each disk in the first moment map of the $^{12}$CO emission (Figure \ref{fig:kinematics}) to check for the presence of a warp, which could be an indicator of the presence of  a (sub)stellar companion. We use the $^{12}$CO data where available, and otherwise $^{13}$CO.  The properties of each moment map are summarized in Table \ref{tbl:kinematics}. 

In 4 disks (IRS~48, HD~142527, MWC~758, J1604-2130) a warp is clearly detected, confirming the results from the literature \citep{Calcino2019,Casassus2013,Boehler2018,Mayama2018}. AB~Aur appears to show non-Keplerian motion as well on larger scales, but this is most likely due to the strong contributions from the spiral arms detected in $^{12}$CO \citep{Tang2017}. For Sz~91 no assessment can be made as the $^{12}$CO emission is optically thin in the inner part of the disk and for DM~Tau, the spectral resolution is very low. For the other disks no warp is detected, but this is possibly due to the low spatial resolution compared to the location of the gap. The detectability of a warp is determined by a combination of signal-to-noise ratio, spectral resolution and spatial resolution compared with the companion orbit radius. Table \ref{tbl:kinematics} lists the relevant parameters for assessing this. We notice that the detected warps have a Beam/$R_{\rm gap}$ value $\lesssim$1.5 and a velocity resolution $\lesssim$0.5 km/s, but overall it remains challenging to determine what specifics set the detectability. 

\begin{table}[!ht]
\centering
\caption{Properties line cubes for first-moment maps}
\label{tbl:kinematics}
\begin{tabular}{llllllllllll}
\hline
Target & Program &  Line & Beam size &Warp & Beam/$R_{gap}$ & $\Delta{v}$ & SNR & Other & Ref & Ref\\
&&&(")&&(km s$^{-1}$)&&&&CO data&other signs\\
\hline
IRS48	&	2013.1.00100.S		&	$^{13}$CO 6-5	&	0.19x0.15	&	Y	&	1.0	&	0.5	&	30	&-& 1&-\\
HD142527	&	2011.0.00465.S		&	$^{12}$CO 3-2	&	0.55x0.33	&	Y	&	1.1	&	0.11	&	60	&shadows& 2&3\\
ABAur	&	2012.1.00303.S		&	$^{12}$CO 3-2	&	0.31x0.19	&	N?	&	0.6	&	0.2	&	43	&mm-disk&4&5\\
MWC758	&	2012.1.00725.S		&	$^{13}$CO 3-2	&	0.22x0.19	&	Y	&	1.5	&	0.11	&	21	&mm-disk&6&5\\
HD135344B	&	2012.1.00158.S		&	$^{13}$CO 3-2	&	0.26x0.21	&	N	&	2.1	&	0.24	&	23	&shadows& 1&7\\
SR21	&	2012.1.00158.S		&	$^{13}$CO 3-2	&	0.23x0.19	&	N	&	4.1	&	0.2	&	25	&NIR CO& 1&8\\
CQTau	&	2013.1.00498.S		&	$^{12}$CO 2-1	&	0.26x0.24	&	N	&	3.2	&	0.3	&	21	&NIR CO&9&10\\
DoAr44	&	2012.1.00158.S		&	$^{13}$CO 3-2	&	0.31x0.29	&	N	&	3.1	&	0.2	&	21	&shadows& 1&11\\
J1604-2130	&	2015.1.00888.S		&	$^{12}$CO 3-2	&	0.23x0.19	&	Y	&	0.8	&	0.21	&	19	&variable,& 12&13,14\\
&&&&&&&&shadows&&\\
LkCa15	&	2012.1.00870.S		&	$^{12}$CO 3-2	&	0.36x0.23	&	N	&	1.9	&	0.21	&	26	&variable&15&16\\
PDS70	&	2017.A.00006.S		&	$^{12}$CO 3-2	&	0.08x0.06	&	N	&	0.4	&	0.42\footnote{Undersampled: see \citet{Keppler2019}}	&	18	&mm-disk& 17&5\\
Sz91	&	2012.1.00761.S		&	$^{12}$CO 3-2	&	0.17x0.13	&	?	&	0.8	&	0.2	&	27	&CO?& 18&19\\
HD169142	&	2013.1.00592.S		&	$^{12}$CO 2-1	&	0.25x0.19	&	N	&	2.1	&	0.16	&	23	&mm-disk& 20 &5\\ 
DMTau	&	2017.1.01460.S		&	$^{12}$CO 2-1	&	0.10x0.10	&	?	&	1.2	&	1.0	&	37	&mm-disk& 21 &5\\
HD163296	&	2016.1.00484.L		&	$^{12}$CO 2-1	&	0.04x0.04	&	N	&	-	&	0.32	&	40	&shadows& 22&23\\
TWHya 	&	2016.1.00629.S		&	$^{12}$CO 3-2	&	0.52x0.41	&	N 	&	-	&	0.05	&	157	&shadows& 24&25\\
\hline
\end{tabular}\\
Refs. 1) \citet{vanderMarel2016-isot}, 2) \citet{Casassus2013}, 3) \citet{Marino2015} 4) Archival data 2012.1.00303, 5) \citet{Francis2020}, 6) \citet{Boehler2017}, 7) \citet{Stolker2016}, 8) \citet{Pontoppidan2008}, 9) Archival data 2013.1.00498, 10) \citet{Chapillon2008}, 11) \citet{Casassus2018}, 12) \citet{Mayama2018}, 13) \citet{Sicilia-Aguilar2020}, 14) \citet{Pinilla2018j1604}, 15) \citet{vanderMarelinprep}, 16) \citet{Alencar2018}, 17) \citet{Keppler2019}, 18) \citet{vanderMarel2018a}, 19) \citet{Tsukagoshi2019}, 20) \citet{Fedele2017}, 21) \citet{Kudo2018}, 22) \citet{Isella2018dsharp}, 23) \citet{Muro-Arena2018}, 24) \citet{Huang2018}, 25) \citet{Debes2017}.
\end{table}

\newpage

\section{Spiral data}
We present the table that was used for creating Figure \ref{fig:lumspirals}. Data were taken from \citet{Garufi2018}, \citet{Francis2020} and \citet{Zhang2018dsharp}. The first 16 targets are from the sample of this study. 

\begin{table}[!ht]
\centering
\caption{Data of properties spiral and non-spiral disks for Figure \ref{fig:lumspirals}.}
\label{tbl:spiraldata} 
\begin{tabular}{llll}
\hline
Target&$L_*$&$R_{\rm dustgapwidth}$&NIR spiral?\\
&($L_{\odot}$&(au)&\\
\hline
IRS48	&	17.8	&	76	&	Y	\\
HD142527	&	9.9	&	200	&	Y	\\
ABAur	&	65	&	170	&	Y	\\
MWC758	&	14	&	50	&	Y	\\
HD135344B	&	6.7	&	51	&	Y	\\
SR21	&	11	&	36	&	Y	\\
DoAr44	&	1.9	&	47	&	N	\\
J1604-2130	&	0.7	&	85	&	N	\\
LkCa15	&	1.3	&	75	&	N	\\
PDS70	&	0.3	&	74	&	N	\\
CQTau	&	10	&	50	&	Y	\\
Sz91	&	0.2	&	94	&	N	\\
HD169142	&	8	&	25	&	M	\\
DMTau	&	0.2	&	25	&	N	\\
HD163296	&	17	&	20	&	N	\\
TWHya	&	0.3	&	5	&	N	\\
HD97048	&	30	&	63	&	M	\\
HD100453	&	6.2	&	30	&	Y	\\
HD100546	&	25	&	27	&	Y	\\
HD142666	&	9	&	16	&	M	\\
AKSco	&	3	&	25	&	Y	\\
GGTau	&	1.6	&	224	&	Y	\\
V4046Sgr	&	0.5	&	31	&	N	\\
LkHalpha330	&	15	&	68	&	Y	\\
GMAur	&	1	&	40	&	M	\\
RXJ1615	&	1.3	&	20	&	N	\\
V1247Ori	&	15	&	64	&	Y	\\
Tcha	&	1.3	&	34	&	M	\\
AS209	&	1.4	&	15	&	M	\\
IMLup	&	2.6	&	5	&	M	\\
RULup	&	1.4	&	8	&	M	\\
RYLup	&	1.9	&	69	&	Y	\\
CSCha	&	1.9	&	37	&	M	\\
J16083070	&	3	&	77	&	M	\\
UXTauA	&	2.5	&	31	&	N	\\
J1852	&	0.6	&	49	&	N	\\
HD143006	&	3.8	&	35	&	Y	\\
\hline
\end{tabular}
\end{table}

\end{document}